\documentclass[aps,prb,twocolumn,groupedaddress]{revtex4-2}

\usepackage{amssymb}
\usepackage{amsmath, physics}
\usepackage{bm}
\usepackage{graphicx}
\usepackage[position=bottom,
justification=RaggedRight,
singlelinecheck=false,
captionskip=-20pt,
labelfont={large,bf},
caption=false
]{subfig}
\usepackage{float}
\usepackage{xcolor}
\usepackage{hyperref}
\usepackage{etoolbox}

\robustify{\subref}

\begin{document}
	
	\title{The topological Faraday effect cannot be observed in a realistic sample}
	
	
	\author{Christian Berger}
	\affiliation{Physikalisches Institut,
		University of W\"urzburg, Am Hubland, D-97074 W\"urzburg, Germany}
	
	\author{Florian Bayer}
	\affiliation{Physikalisches Institut,
		University of W\"urzburg, Am Hubland, D-97074 W\"urzburg, Germany}
	
	\author{Laurens W. Molenkamp}
	\affiliation{Physikalisches Institut,
		University of W\"urzburg, Am Hubland, D-97074 W\"urzburg, Germany}
	
	\author{Tobias Kiessling}
	\affiliation{Physikalisches Institut,
		University of W\"urzburg, Am Hubland, D-97074 W\"urzburg, Germany}
	\email{tobias.kiessling@physik.uni-wu
		erzburg.de}
	
	
	\date{\today}
	
	\begin{abstract}
		A striking feature of 3 dimensional (3D) topological insulators (TIs) is the theoretically expected topological magneto-electric (TME) effect, which gives rise to additional terms in Maxwell's laws of electromagnetism with an universal quantized coefficient proportional to half-integer multiples of the fine structure constant $\alpha$. In an ideal scenario one therefore expects also quantized contributions in the magneto-optical response of TIs. We review this premise by taking into account the trivial dielectric background of the TI bulk and potential host substrates, and the often present contribution of itinerant bulk carriers. We show that (i) one obtains a non-universal magneto-optical response whenever there is impedance mismatch between different layers and (ii) that the detectable signals due to the TME rapidly approach vanishingly small values as the impedance mismatch is detuned from zero. We demonstrate that it is methodologically impossible to deduce the existence of a TME exclusively from an optical experiment in the thin film limit of 3D TIs at high magnetic fields.
	\end{abstract}
	
	\maketitle

	\section{Introduction}
	
	The hallmark feature of three dimensional (3D) topological insulators (TIs) is the existence of a quantized surface conductance that arises from the topological bulk properties and of the solid state system \cite{2010Hasan, 2011Qi, 2005KaneZ2, 2006BHZ, 2009Roy, 2007FuInversion, 2007Fu}. It is quantized in units of $(\frac{1}{2}+n)e^2/h$, in which $e$ denotes the electron charge, $h$ the Planck constant and $n=(0,1,2,\dots)$. Another key characteristic of the TI surface states is the coupling of the spin and momentum degrees of freedom of TI surface conducting electrons \cite{2005KaneQSH, 2006Bernevig}. In the theoretical limit of a perfectly insulating bulk, this coupling gives rise to the existence of a bulk linear magneto-electric effect in the form of an $\bm{E}\cdot\bm{B}$ term, which constitutes the topological magneto-electric effect (TME) \cite{2008Qi, 2009Vanderbilt}. The latter is sometimes conceived as an independent effect, but is really just a way of describing the 3D TI conductance in terms of a bulk magneto-electric material, provided it is geometrically possible to define a magneto-electric polarization~\cite{2013Bernevig}. It is therefore instantly clear, that if the surface conduction is quantized so must be the TME and vice versa.
	
	The experimental reality is much less clear. Any realistic TI sample will have a finite amount of bulk carriers and even for very small amounts ($<10^{16}$ cm$^{-3}$) it is not obvious up to which point the considerations that yield the $\bm{E}\cdot\bm{B}$ term hold in practice. A further practical limitation arises from the fact that the cleanest available samples are thin films \cite{2011Bruene, Armitage_samples}, for which it is difficult to experimentally access the bulk portion of the 3D TI in which the TME resides. The question therefore is: How can the presence of a quantized TME unambiguously be demonstrated in an experiment?
	
	One very early proposal was to employ magneto-optical polarimetry to this end, in particular, polar Faraday/Kerr rotation experiments. The presence of the $\bm{E}\cdot\bm{B}$ term will modify the continuity conditions at the interfaces/surfaces and translate the quantized surface conductance into a quantized and therefore clearly distinguishable Faraday/Kerr response \cite{2008Qi}. The key prediction was that the Faraday response should be quantized in units of integral multiples of the fine structure constant $\alpha$, and this \textit{universal topological Faraday Effect (TFE)} was thought to be a signature feature of 3D TIs \cite{2008Qi}. Later theoretical work refined this idea in terms of the available material systems, for which due to the relevant energy scales the experiment had to be performed in the (far) infrared spectral region (we will elaborate on this aspect below) \cite{2010MacDonald}. The latter constraint is a hard one, because the long wavelenghts involved prevent a direct measurement of bulk properties by optical techniques for the thin films. In particular, they inhibit the separation of the optical response of the two surfaces in the time domain.
	
	Nonetheless, claims on the experimental observation of the TFE followed soon \cite{2016Armitage,2016Pimenov,2016Okada}. In these works Faraday rotations that either extrapolate to or are to experimental accuracy within $\alpha$ were reported. These results quickly gave rise to new questions. Beenakker pointed out that in the stratified slab geometries employed in these works the TFE contributions from the two surfaces should exactly cancel \cite{2016Beenakker}. A general question is further, how unique Faraday rotations close to $\alpha$ really are and how one can clearly assign the physical origin of such a signal. And finally, the question arises, whether or not it is at all methodically possible, to unambiguously deduce the existence of a TME term exclusively from the polarimetry response in the thin film limit.
	
	We address these questions by calculating the magneto-optical response of a homogeneous film that hosts a TME in the semiclassical limit, i. e. we make no particular microscopic assumptions beyond the existence of a 3D strong TI \cite{2013Bernevig} and describe the interaction with the electromagnetic wave in a classical fashion. We explicitly assume a description that builds on the existence of an $\bm{E}\cdot\bm{B}$ term rather than a picture that starts from the surface conductance. This is the more natural view when the TME is the subject of the experimental study and the two perspectives are interchangeable in the clean theoretical limit.  The resulting framework is valid for any linear magneto-electric material and useful beyond the scope of the TME, but here we confine the discussion to the case of the 3D TI.
	
	The paper is organized as follows: We first derive and discuss the response in the (idealized) clean dielectric limit. Second, we study the scenario in which residual bulk carriers coexist with TI surface states. Third, we discuss the connection to the conceptionally close but fundamentally different AC response of a quantum Hall system. We finally demonstrate that the observables of a Faraday/Kerr experiment do not allow for an unambiguous assignment of the presence of a TME in the thin film limit.
	
	\section{Optical polarization rotation in 3D TIs}
	
	Polarimetry experiments measure the polarization of the electromagnetic wave when being reflected from (Kerr geometry) or transmitted through (Faraday geometry) an interface at which the impedance properties change. The impedance change can generally be induced by either changes in the permittivity and/or the permeability. For magneto-electrically active materials there is further a direct coupling between $\bm{E}$ and $\bm{B}$ components, which actually means that the electric field induces a magnetization and the magnetic induction a polarization \cite{2005Fiebig}. For the linear magneto-electric (ME) effect this coupling results in terms proportional to $\bm{E}\cdot\bm{B}$ in the Lagrangian of the system\cite{2009Hehl}.
	
	A formally similar term arises in the context of the hypothetical cosmological axion, which is why the magneto-electric term in topological systems has been coined axion term or sometimes $\theta$-term, as the coupling coefficient is often labelled such. This labelling is rather misleading, because the cosmological axion has a very different physical meaning than the TME and in fact any linear magneto-electric contribution will have a formal resemblance to the axion term \cite{2009Hehl}, without actually describing the same physics.
	
	For the remainder of this work we will call the $\bm{E}\cdot\bm{B}$ term simply the TME and take the corresponding response as explicit terms that we add to the dielectric displacement and the magnetic induction. We split the permittivity into a dielectric contribution into which we lump the total response of all non-itinerant charge carriers of the system (the classic dielectric response), and the conductivity contribution arising from free carriers.
	
	\subsection{The clean dielectric limit} \label{ssec:clean_dielectric_limit}
	
	We first consider the idealized situation in which there are no free carriers in the system at all, which we refer to as the clean dielectric limit. In the most general case the optical polarization state of a electro-magnetic (EM) wave is elliptical, which is described by two mutually orthogonal linear polarization components $E_\bot$ and $E_\|$ that are shifted by some phase in real space. We first define the complex quantity $\rho$ as 
	\begin{equation}
		\rho=\frac{E_\|}{E_\bot}.
	\end{equation}
	For Faraday/Kerr measurements one usually starts out with a well defined linearly polarized wave, which we take to be polarized along $E_\bot$. Any induced rotation of the plane of polarization will then be given by the in-phase contribution of $\rho$. We hence define the rotation angle $\theta$ as
	\begin{equation}
		\theta=\Re{\arctan(\rho)}. \label{eqTheory:rotation_definition}
	\end{equation}
	The ellipticity $\phi$ is the angle provided by the ratio of the out of phase component of $\rho$ and is given by~\cite{2012Armitage}
	\begin{equation}
		\phi=\frac{1}{2}\arcsin(\tanh(2\Im{\arctan(\rho)})). \label{eqTheory:ellipticity_definition}
	\end{equation}
	
	We then use the Jones formalism to calculate the transmitted electric field components. The effect of any interface on the polarization state can be described by a 2-by-2 transmission matrix
	\begin{align}
		\underline{T}&=\pmqty{t_{ss} & t_{sp} \\ t_{ps} & t_{pp}}
	\end{align}
	Upon transmission through a multi-layer stack one generally has to take into account the optical thickness of layers, which gives rise to the well known Fabry-P\'erot patterns. Observing the TME requires gapping out the Dirac cone, which is commonly done with a magnetic field. The Dirac gap needed, however, is typically only a few $\text{meV}$ in energy~\cite{2011Qi}. Because of this and also to avoid photodoping by photocarrier excitation to energetically higher bulk bands, Faraday/Kerr experiments on the TME have been performed in the far infrared (FIR) spectral region.
	
	On the other hand, typical sample thicknesses are well below 100~$\text{nm}$ and often only a few monolayers, which is at least three orders of magnitude thinner than the optical wavelength in the FIR regime. We therefore neglect finite thickness contributions for the remainder of this work. Such contributions are in principle straightforward to add at a later stage and will have no impact on the general results stemming from the TME, which only contributes at the interfaces.
	
	In the thin film limit the resulting electric field vector $\bm{E}_t$ after interaction with $n$ interfaces is then given by:
	\begin{align}
		\pmqty{E_{t,\bot} \\ E_{t,\|}}&=\prod_{k=1}^{n}\underline{T}_{n-k+1} \pmqty{E_{i,\bot} \\ E_{i,\|}},
	\end{align}
	where the index of $\underline{T}$ denotes the interface number.
	
	To derive the elements of $\underline{T}$, we start with the continuity conditions for EM waves at an interface between two materials $a$, $b$:
	\begin{align}
		\begin{aligned}
			\bm{n}\cdot\bm{D}&=0  \qquad&\bm{n}\cdot\bm{B}=0\\
			\bm{n}\times\bm{E}&=0  \qquad&\bm{n}\times\bm{H}=0 \label{eqTheory:continuity_conditions}
		\end{aligned}					
	\end{align}
	Due to the TME, there are collinear electric and magnetic fields arising which alter the usual definitions of the electric displacement field $\bm{D}$ and the magnetic field $\bm{H}$~\cite{2008Qi}:
	
	\begin{align}
		\begin{aligned}
			\bm{D}=\varepsilon_r\varepsilon_0\bm{E} - 2P_3\alpha\sqrt{\frac{\varepsilon_0}{\mu_0}}\bm{B}\\
			\bm{H}=\frac{1}{\mu_r\mu_0}\bm{B} + 2P_3\alpha\sqrt{\frac{\varepsilon_0}{\mu_0}}\bm{E} \label{eqTheory:TME_DH}
		\end{aligned}
	\end{align}
	with relative permittivity $\varepsilon_r$, relative permeability $\mu_r$, vacuum permittivity $\varepsilon_0$, vacuum permeability $\mu_0$, the TME polarization $P_3$, and the fine-structure constant $\alpha$.\\
	From Eqs. \eqref{eqTheory:continuity_conditions} and \eqref{eqTheory:TME_DH}, we derive the relations between incident and transmitted field components, from which we obtain the transmission matrix elements by comparing coefficients (the derivation is given in detail in Appendix A). For normal incidence this yields
	
	\begin{align}
		\Delta&=\left(\sqrt{\frac{\varepsilon_{r,a}}{\mu_{r,a}}}+\sqrt{\frac{\varepsilon_{r,b}}{\mu_{r,b}}}\right)^2+\left(2\left(P_{3,b}-P_{3,a}\right)\alpha\right)^2 \label{eqTheory:delta_epsnoN}\\
		t_{ss}&=t_{pp}=\frac{2}{\Delta}\sqrt{\frac{\varepsilon_{r,a}}{\mu_{r,a}}}\left(\sqrt{\frac{\varepsilon_{r,a}}{\mu_{r,a}}}+\sqrt{\frac{\varepsilon_{r,b}}{\mu_{r,b}}}\right) \label{eqTheory:tss_epsnoN}\\
		t_{sp}&=-t_{ps}=-\frac{4}{\Delta}\sqrt{\frac{\varepsilon_{r,a}}{\mu_{r,a}}}\left(P_{3,b}-P_{3,a}\right)\alpha \label{eqTheory:tsp_epsnoN}
	\end{align}
	
	In a similar fashion, one can derive the components of the reflection matrix ($\Delta$ same as for transmission) and obtain for normal incidence:
	
	\begin{align}
		r_{ss}&=-r_{pp}=\frac{1}{\Delta}\left[\frac{\varepsilon_{r,a}}{\mu_{r,a}}-\frac{\varepsilon_{r,b}}{\mu_{r,b}}-\left(2\left(P_{3,b}-P_{3,a}\right)\alpha\right)^2\right] \label{eqTheory:rss_epsnoN}\\
		r_{sp}&=r_{ps}=-\frac{4}{\Delta}\sqrt{\frac{\varepsilon_{r,a}}{\mu_{r,a}}}\left(P_{3,b}-P_{3,a}\right)\alpha \label{eqTheory:rsp_epsnoN}
	\end{align}
	
	We next study the polarization response for a linearly $s$-polarized EM wave propagating through a single interface of materials $a$ and $b$ at normal incidence.
	For the transmitted wave we obtain
	\begin{align}
		\pmqty{E_{t,\bot}\\E_{t,\|}}=\pmqty{t_{ss} & t_{sp} \\ -t_{sp} & t_{ss}}\pmqty{1 \\ 0}=\pmqty{t_{ss} \\ -t_{sp}}
	\end{align}
	
	and the complex Faraday rotation is defined by (analogous for Kerr rotation $\theta_K$)
	
	\begin{align}
		\theta_F = \arctan(-\frac{t_{sp}}{t_{ss}}).
	\end{align}
	
	Up to this point the dielectric functions are real and accordingly only a rotation but no ellipticity can be induced.
	Let material $a$ be topologically trivial and $b$ topologically non-trivial, then $P_{3,a}=0$ and $P_{3,b}=1/2\;$(mod 1). For the resulting Faraday/Kerr rotation we finally obtain
	
	\begin{align}
		\theta_{F,1} &= \arctan(\frac{4\sqrt{\frac{\varepsilon_{r,a}}{\mu_{r,a}}}\left(P_{3,b}-P_{3,a}\right)\alpha}{2\sqrt{\frac{\varepsilon_{r,a}}{\mu_{r,a}}}\left(\sqrt{\frac{\varepsilon_{r,a}}{\mu_{r,a}}}+\sqrt{\frac{\varepsilon_{r,b}}{\mu_{r,b}}}\right)})\label{eqTheory:FaradayEps_oneInterface}\\
		&=\arctan(\frac{1}{\sqrt{\frac{\varepsilon_{r,a}}{\mu_{r,a}}}+\sqrt{\frac{\varepsilon_{r,b}}{\mu_{r,b}}}}\;\alpha) \label{eqTheory:FaradayEps_oneInterface_simplified}\\
		\theta_{K,1} &=			\arctan(\frac{-2\sqrt{\frac{\varepsilon_{r,a}}{\mu_{r,a}}}\,\alpha}{\frac{\varepsilon_{r,a}}{\mu_{r,a}}-\frac{\varepsilon_{r,b}}{\mu_{r,b}}-\alpha^2})\label{eqTheory:KerrEps_oneInterface}
	\end{align}
	
	\begin{figure}[t]
		\includegraphics[width=\linewidth]{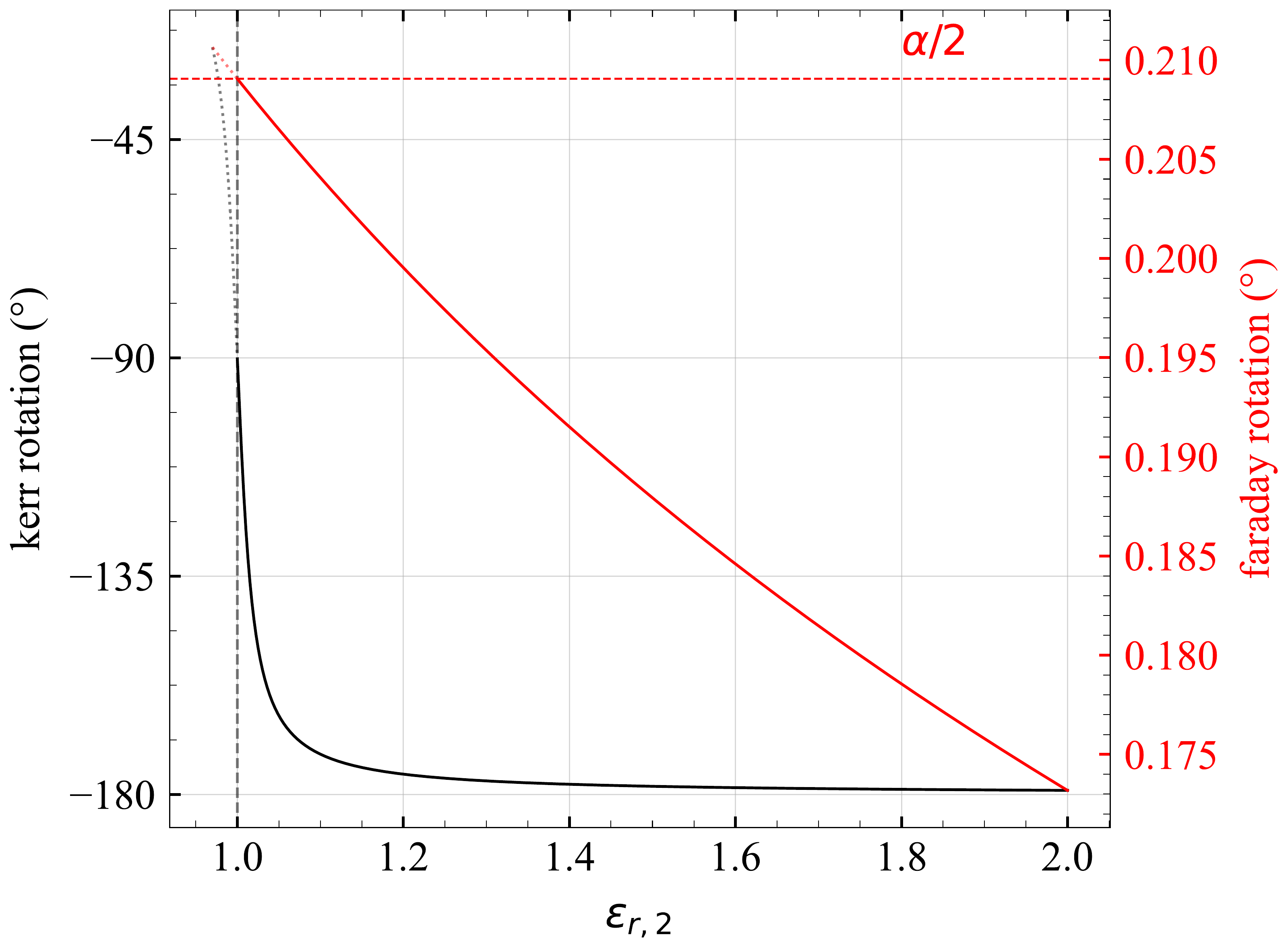}
		\caption{Kerr and Faraday rotation after interaction with a single interface in dependence of the permittivity of the second material (while $\varepsilon_{r,1}=1$). The dotted lines indicate the region of parameter space that cannot be accessed with physically reasonable material values.}
		\label{figTheory:rotation_vs_permittivity}
	\end{figure}
	
	
	Let us now examine Eqs.~\ref{eqTheory:FaradayEps_oneInterface_simplified} and \ref{eqTheory:KerrEps_oneInterface} in detail. We immediately recognize one important result: \textit{There is no universal topological Faraday or Kerr effect}. This refinement of the initial theoretical predication was also stated by the original authors~\cite{2011Qi}, but obviously this message has not been widely received and even recent experimental publications keep repeating claims on the existence of a universal Faraday response~\cite{2022Tokura}. The optical polarization response heavily depends on the impedance mismatch between the layers, which in case of non-magnetic layers reduces to the dielectric mismatch. We stress that this is a feature of the TFE itself. Up to this point we deal with isotropic materials without free carriers, so without the TME contribution there is no Faraday/Kerr rotation at all (at normal incidence). This is also immediately verified upon inspection of Eqs.~\ref{eqTheory:tsp_epsnoN} and \ref{eqTheory:rsp_epsnoN}, which yield zero for $t_{sp}$ and $r_{sp}$ if $P_3$ remains zero and accordingly the Faraday/Kerr rotation vanishes.
	
	For a quantitative discussion we plot in Fig.~\ref{figTheory:rotation_vs_permittivity} the magnitude of the topological Faraday/Kerr angle against the dielectric mismatch of the layers with fixed $\varepsilon_{r,1}=1$ (vacuum). In the case of dielectrically matched layers we restore the initial prediction of the TFE, which is $\approx90^{\circ}$ for the Kerr angle and a Faraday angle equal to $\alpha/2$ for a single interface \cite{2008Qi} (only when $\varepsilon_{r,1/2}=1$). 
	
	For non-zero dielectric mismatch the rotation angles \textit{continuously} detune to different values. Within reasonable physical limits ($\varepsilon_{r,1}/\varepsilon_{r,2}\leq1$) the Faraday angle can only be $\leq\alpha/2$ and decreases with increasing dielectric constants. In reflection geometry this produces systematically smaller Kerr angles values, which rapidly approach zero (the $-180^{\circ}$ values the well known phase jump upon reflection off media with higher optical density). The physical reason for this behaviour is readily understood. For the dielectrically matched case there is no regular reflection~\footnote{More accurately, the sheer existence of a magnetoelectric effect already introduces an impedance mismatch between the two materials leading to a regular reflection even with matching permittivities and permeabilities. In case of the TME, this contribution is scaling with $\alpha^2$ \eqref{eqTheory:rss_epsnoN} (while the rotation-driving contribution scales with $\alpha$ \eqref{eqTheory:rsp_epsnoN}) and therefore is very small.}. The TME term couples an $E_{\|}$ component to the transmitted $E_\bot$, which gives rise to Faraday rotation, the magnitude of which is set by the coupling constant of $E_\bot$ and $E_{\|}$. To maintain the continuity of the electric field at the interface a component of $-E_{\|}$ needs to be simultaneously reflected. Since there is no other reflection, this results in a Kerr angle of $-90^{\circ}$.
	
	However, the \textit{magnitude} of the reflected $-E_{\|}$ is rather small. With dielectric mismatch there is a component of $E_\bot$ reflected, which rapidly increases with increasing dielectric mismatch, effectively rotating the plane of polarization back to the incident polarization state. With part of the incident $E_\bot$ now reflected, the magnitude of the transmitted portion of $E_\bot$ decreases. Since the transmission factors $t_{ss}$ and $t_{pp}$ scale with $\varepsilon_{r,a}$, while $t_{sp}$ and $t_{ps}$ scale with $\sqrt{\varepsilon_{r,a}}$, this results in a continuously decreasing rotation angle with increasing $\varepsilon_{r,a}$ and vice versa, obviously impacting the Kerr angle.
	
	In summary, while the TME is quantized in units of $\alpha$, the TFE is NOT quantized at all and in fact can take different values depending on the exact dielectric mismatch.
	
\begin{figure}[t]
		\includegraphics[width=\linewidth]{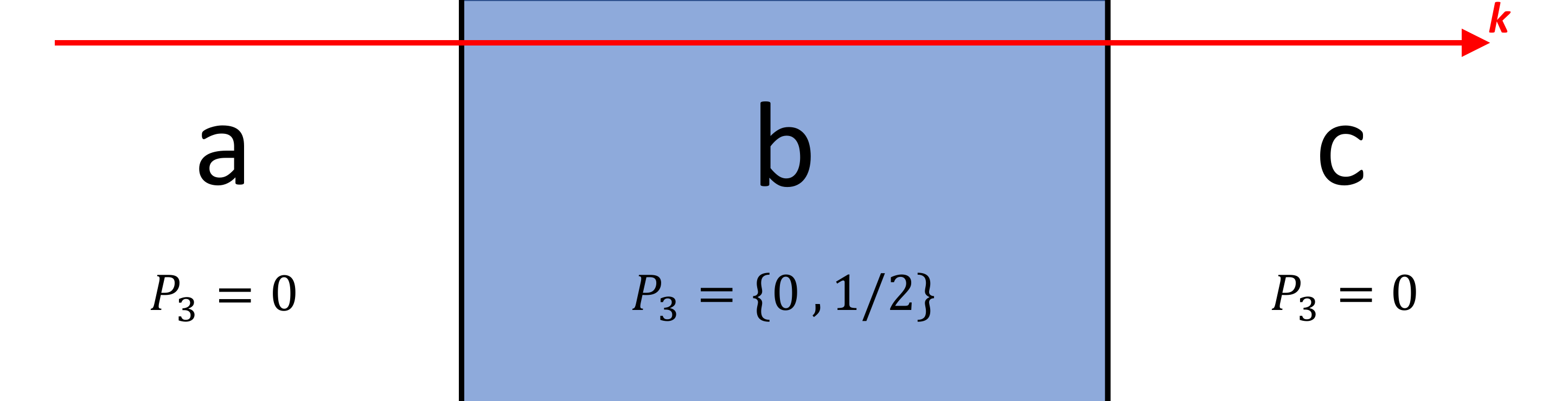}\\\vspace{10pt}
		\includegraphics[width=\linewidth]{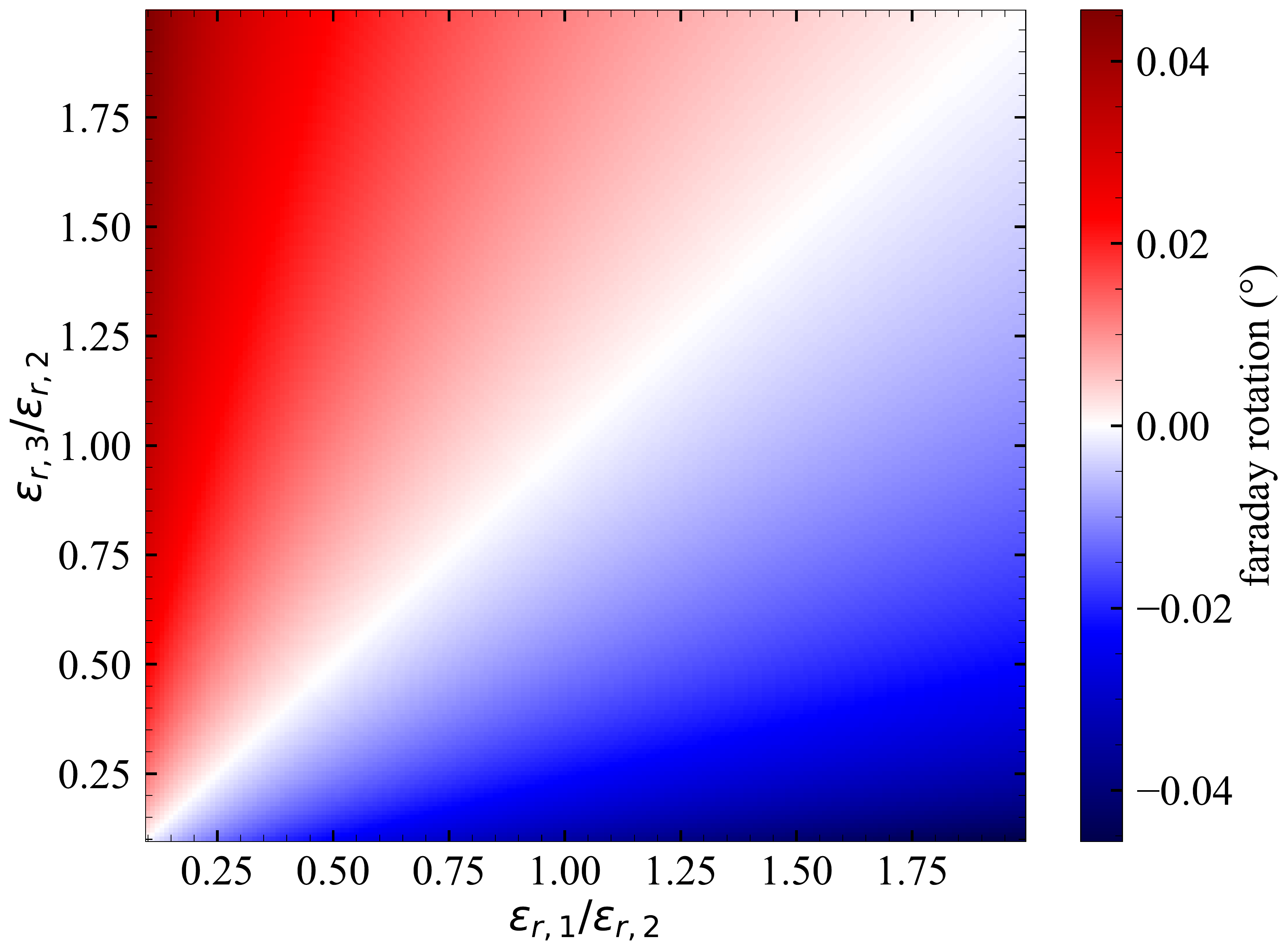}
		\caption{\textbf{top}: General layer-stack for two interfaces. The contribution from TME is indicated by the "magnetoelectric polarization" $P_3$, which can only be $0$ (no TME-contribution) or $1/2$ (TME-contribution) \cite{2008Qi}. \textbf{bottom}: Faraday rotation through two interfaces in slab geometry for different combinations of permittivities of all three layers. Layers one and three are taken to be topologically trivial ($P_3=0$), layer two hosts the TME ($P_3=\frac{1}{2}$) with its permittivity set to 10.}
		\label{figTheory:rotation_dependency_permittivity_mismatch}
	\end{figure}

	
	After having discussed the fundamentals, we shall now assess the experimentally more relevant transmission through a slab geometry as depicted in Fig.~\ref{figTheory:rotation_dependency_permittivity_mismatch}. 
	We here follow a more general approach to the total rotation after transmission through two interfaces:
	\begin{align*}
		\pmqty{E_{t,\bot}\\E_{t,\|}}&=\underline{T_b}\,\underline{T_a}\,\pmqty{E_{i,\bot}\\E_{i,\|}}\\
		&=\pmqty{(t_{ss,a}t_{ss,b}-t_{sp,a}t_{sp,b})\,E_{i,\bot}\\+(t_{ss,a}\,t_{sp,b}+t_{sp,a}\,t_{ss,b})\,E_{i,\|}\\\\
			-(t_{ss,a}\,t_{sp,b}+t_{sp,a}\,t_{ss,b})\,E_{i,\bot}\\+(t_{ss,a}t_{ss,b}-t_{sp,a}t_{sp,b})\,E_{i,\|}}
	\end{align*}
	In order to preserve the initial rotation angle
	\begin{align}
		t_{ss,a}\,t_{sp,b}+t_{sp,a}\,t_{ss,b} \overset{!}{=} 0 \label{eq:ConditionNoRotation}
	\end{align}
	must hold.\\
	For a symmetric configuration, the transmission coefficients can be represented as
	\begin{align*}
		\Delta &= \Delta_a = \Delta_b\\
		t_{ss,a}&=\frac{2}{\Delta}\sqrt{\frac{\varepsilon_{r,a}}{\mu_{r,a}}}\left(\sqrt{\frac{\varepsilon_{r,a}}{\mu_{r,a}}}+\sqrt{\frac{\varepsilon_{r,b}}{\mu_{r,b}}}\right)\\
		t_{sp,a}&=-\frac{4}{\Delta}\sqrt{\frac{\varepsilon_{r,a}}{\mu_{r,a}}}\left(P_{3,b}-P_{3,a}\right)\alpha\\
		t_{ss,b}&=\frac{2}{\Delta}\sqrt{\frac{\varepsilon_{r,b}}{\mu_{r,b}}}\left(\sqrt{\frac{\varepsilon_{r,a}}{\mu_{r,a}}}+\sqrt{\frac{\varepsilon_{r,b}}{\mu_{r,b}}}\right)\\
		t_{sp,b}&=\frac{4}{\Delta}\sqrt{\frac{\varepsilon_{r,b}}{\mu_{r,b}}}\left(P_{3,b}-P_{3,a}\right)\alpha
	\end{align*}
	which leads to
	\begin{align*}
		t_{ss,a}\,t_{sp,b} =& \frac{4}{\Delta^2}\,\sqrt{\frac{\varepsilon_{r,a}\,\varepsilon_{r,b}}{\mu_{r,a} \,\mu_{r,b}}}\,\left(\sqrt{\frac{\varepsilon_{r,a}}{\mu_{r,a}}}+\sqrt{\frac{\varepsilon_{r,b}}{\mu_{r,b}}}\right)\\
		&\cdot(P_{3,b}-P_{3,a})\,\alpha\\
		t_{sp,a}\,t_{ss,b} =& -\,t_{ss,a}\,t_{sp,b}
	\end{align*}
	and Eq.\eqref{eq:ConditionNoRotation} is always valid, which reproduces the Beenakker argument that the TME contributions should cancel~\cite{2016Beenakker}.

	For non-identical impedance mismatch at the two interfaces we obtain, again, a continuous evolution of the Faraday rotation away from zero values, as plotted in Fig.~\ref{figTheory:rotation_dependency_permittivity_mismatch} for $\varepsilon_{r,2}=10$. The general trend is an increase of the Faraday rotation angle with increasing impedance mismatch, capping at the expected value for a single interface when $\varepsilon_{r,1/3}$ becomes very large. The sign of the rotation angle is entirely determined by the mismatch ratios. As main result, we again obtain that there is no universal topologically defined quantized Faraday rotation.\\
	For the Kerr rotation, only the interface on which the wave is reflected matters in the thin film approximation and no further insight is gained. Any additional interfaces do not add to the absolute value of rotation, since any individual interface, which is passed twice but in different directions, can be treated as two interfaces with symmetrical layer stack, for which the total rotation vanishes.
	
	As a closing remark, we point out that while we restrict the discussion in this manuscript to non-magnetic systems, the obtained results implicate that magneto-optical detection of a potential TME contribution in 3D quantum anomalous spin Hall systems is outright impractical. Set apart from the fact that any TME contribution is going to be vanishingly small against a ferromagnetic magnetization to begin with, the resulting impedance mismatch between the ferromagnetic and non-magnetic layers is enormous. From the above it is then immediately clear that the magneto-optical response is likewise going to be dominated by the impedance mismatch and any TFE signal is again not quantized and extremely small against the regular ferromagnetic magneto-optical contribution, likely below any available practical detection threshold.
	
	\subsection{Residual bulk carrier contribution}
	Up to this point the impact of itinerant carriers on the magneto-optical response has been neglected. This is a common approximation in the bulk of the theoretical literature~\cite{2008Qi,2010MacDonald}, but does not very well resemble the experimental reality. All experimentally available TI systems host a finite amount of bulk carriers. Even excellent materials require the control of a gate~\cite{2007Koenig} or need to be probed at mK temperatures~\cite{2011Bruene} to not be dominated in their physical response by the residual bulk carriers. For the magneto-optical response of the TME we have found the impedance mismatch of the layers to be decisive. As a general statement, the influence of itinerant carriers on the continuity conditions of the fields at the interface is striking even for very low carrier concentrations (and completely dominates already at mediocre concentrations). This is particularly true in the thin film limit, for which one often neglects the dielectric background of the layer altogether~\cite{2016Armitage}. It is therefore likely that any realistic (i.e. quantitative) model has to take itinerant carriers into account. 
	
	To model the impact of residual bulk carriers we follow the well established literature~\cite{1970Furdyna} and add the TME terms to the formalism. For the sake of clarity we briefly summarize the derivation instead of just giving the result. The current density $\bm{j}$ connects to the electric field $\bm{E}$ via Ohm's law
	\begin{equation}
		\bm{j} \equiv \underline{\sigma} \bm{E} \label{eqTheory:Ohm'sLaw}
	\end{equation}
	through the conductivity tensor $\underline{\sigma}$. The tensor character of the conductivity arises from the anisotropy that is induced by the presence of a static magnetic field $\bm{B}_0$ needed to break time reversal symmetry, a requirement for the TME to be observable~\footnote{If this is achieved by other means, the situation simplifies, but since all existing experimental work employed magnetic fields to break time reversal symmetry, we consider the more general scenario.}. We assume isotropic media and align $\bm{B}_0$ along the $z$-axis, which we further take to be normal to our sample (i. e. Faraday geometry). The conductivity tensor $\underline{\sigma}$ then takes the form
	\begin{align}
		\underline{\sigma}&=\pmqty{\sigma_{xx} & \sigma_{xy} & 0 \\ -\sigma_{xy} & \sigma_{xx} & 0 \\ 0 & 0 & \sigma_{zz}} \label{eqTheory:ConductivityTensor}
	\end{align}
	without loss of generality. Using this, we define the generalized dielectric tensor $\underline{\varepsilon}$
	\begin{align}
		\underline{\varepsilon}=\varepsilon_{r}\;\mathbb{I}+\frac{\text{i}}{\omega\varepsilon_0}\;\underline{\sigma} \label{eqTheory:complexEpsilon},
	\end{align}
	with relative permittivity $\varepsilon_r$ that contains the contribution of all non-itinerant carriers, unity matrix $\mathbb{I}$, frequency of light $\omega$ and vacuum permittivity $\varepsilon_0$.
	In the given geometry the dielectric tensor then takes the explicit form
	\begin{align}
		\underline{\varepsilon}=\pmqty{\varepsilon_{xx}&\varepsilon_{xy}&0 \\ -\varepsilon_{xy}&\varepsilon_{xx}&0 \\ 0&0&\varepsilon_{zz}}\label{eqTheory:DielectricTensor}
	\end{align}
	
	which is commonly labelled as gyrotropic dielectric tensor and for which the components take the explicit form
	
	\begin{align}
		\varepsilon_{xx}&=\varepsilon_r + \frac{\text{i}}{\omega\varepsilon_0}\;\sigma_{xx}\\
		\varepsilon_{xy}&=\frac{\text{i}}{\omega\varepsilon_0}\;\sigma_{xy}\\
		\varepsilon_{zz}&=\varepsilon_r + \frac{\text{i}}{\omega\varepsilon_0}\;\sigma_{zz}\label{eqTheory:DielectricTensorComponents}		
	\end{align}

	From Eqs.~\eqref{eqTheory:delta_epsnoN}-\eqref{eqTheory:rsp_epsnoN} it is evident that the square root of the dielectric function is required for the derivation of the transmission/reflection matrix. We hence need to calculate the corresponding ``square root matrix'', which satisfies the equation
	
	\begin{align}
		\underline{\gamma}\,^2&=\underline{\varepsilon}
	\end{align}
	
	For our choice of coordinates the solution to this equation is
	
	\begin{align} 
		\underline{\gamma}&=\pmqty{\gamma_{xx}&\gamma_{xy}&0\\ -\gamma_{xy}&\gamma_{xx}&0 \\ 0&0&\gamma_{zz}}
	\end{align}
	
	with the components
	
	\begin{align}
		\gamma_{xx}&=\frac{1}{2}\left(\sqrt{\varepsilon_{xx}-\text{i}\varepsilon_{xy}}+\sqrt{\varepsilon_{xx}+\text{i}\varepsilon_{xy}}\right)\\
		\gamma_{xy}&=\frac{\text{i}}{2}\left(\sqrt{\varepsilon_{xx}-\text{i}\varepsilon_{xy}}-\sqrt{\varepsilon_{xx}+\text{i}\varepsilon_{xy}}\right)\\
		\gamma_{zz}&=\sqrt{\varepsilon_{zz}}	\label{eq:gamma_component_definition}
	\end{align}
	
	We then perform the same derivation as in \ref{ssec:clean_dielectric_limit}, but with the complex dielectric function matrix. The derivation is shown in detail in Appendix~\ref{appsec:derivation_transmission_matrix} and finally yields for the resulting transmission and reflection matrix elements:
	\begin{widetext}
		\begin{align}
			\Delta=&\left(\sqrt{\frac{1}{\mu_a}}\;\gamma_{xx,a}+\sqrt{\frac{1}{\mu_b}}\;\gamma_{xx,b}\right)^2+\left(\sqrt{\frac{1}{\mu_a}}\;\gamma_{xy,a}+\sqrt{\frac{1}{\mu_b}}\;\gamma_{xy,b}+2\left(P_{3,b}-P_{3,a}\right)\alpha\right)^2  \label{eqTheory:TransmittivityWithCarriers}\\
			t_{ss}=t_{pp}=&\frac{2}{\Delta}\;\sqrt{\frac{1}{\mu_a}}\left[\sqrt{\frac{1}{\mu_a}}\;\left(\gamma_{xy,a}^2+\gamma_{xx,a}^2\right)+\sqrt{\frac{1}{\mu_b}}\;\left(\gamma_{xy,a}\gamma_{xy,b}+\gamma_{xx,a}\gamma_{xx,b}\right)+2\left(P_{3,b}-P_{3,a}\right) \gamma_{xy,a}\;\alpha\right]\\
			t_{sp}=-t_{ps}=&\frac{2}{\Delta}\;\sqrt{\frac{1}{\mu_a}}\;\left(\sqrt{\frac{1}{\mu_b}}\left(\gamma_{xx,b}\gamma_{xy,a}-\gamma_{xx,a}\gamma_{xy,b}\right)-2\left(P_{3,b}-P_{3,a}\right)\gamma_{xx,a}\;\alpha\right)\\
			r_{ss}=-r_{pp}=&\frac{1}{\Delta}\left[\left(\sqrt{\frac{1}{\mu_a}}\;\gamma_{xy,a}-\sqrt{\frac{1}{\mu_b}}\;\gamma_{xy,b}-2\left(P_{3,b}-P_{3,a}\right)\alpha\right)
			\left(\sqrt{\frac{1}{\mu_a}}\;\gamma_{xy,a}+\sqrt{\frac{1}{\mu_b}}\;\gamma_{xy,b}+2\left(P_{3,b}-P_{3,a}\right)\alpha\right)\right.\nonumber\\
			&\hphantom{\frac{1}{\Delta}\left[\right.}\left.+\frac{1}{\mu_a}\;\gamma_{xx,a}^2-\frac{1}{\mu_b}\;\gamma_{xx,b}^2\right]\\
			r_{sp}=r_{ps}=&\frac{2}{\Delta}\;\sqrt{\frac{1}{\mu_a}}\;\left(\sqrt{\frac{1}{\mu_b}}\left(\gamma_{xx,b}\gamma_{xy,a}-\gamma_{xx,a}\gamma_{xy,b}\right)-2\left(P_{3,b}-P_{3,a}\right)\gamma_{xx,a}\;\alpha\right) \label{eqTheory:ReflectivityWithCarriers}
		\end{align}	
	\end{widetext}

	For a more quantitative discussion on the impact of residual bulk carriers we need to model the free carrier conductivity. A widely applied approach for the analysis of the infrared spectral response of free carriers is the Drude model. For the degenerate case (i.~e. electro-chemical potential resides in the conduction band) the Drude model actually reproduces the analytical form obtained from the Boltzmann transport equation in relaxation time approximation~\cite{1970Furdyna}, and will therefore generally be a good starting point for experiments on the BiSe material family. It is further rather successful upon describing the sub-THz AC transport response in the diffusive regime for wide class of materials~\cite{1970Furdyna}. We emphasize that our goal is not to find a quantitative description for the most general case but rather to establish general trends. Conductivities obtained from more powerful models can always be inserted into Eqs.~~\eqref{eqTheory:TransmittivityWithCarriers}~-~\eqref{eqTheory:ReflectivityWithCarriers}.
	
	In the Drude model, the components of $\underline{\sigma}$ take the following form: 
	
	\begin{align}
		\sigma_{xx}&=\frac{ne^2\tau^*}{m}\frac{1}{1+\left(\omega_c\tau^*\right)^2}\\
		\sigma_{xy}&=\frac{ne^2\tau^*}{m}\frac{\omega_c\tau^*}{1+\left(\omega_c\tau^*\right)^2}\\
		\sigma_{zz}&=\frac{ne^2\tau^*}{m} \label{eqTheory:defConductivityMatrix}
	\end{align}
	
	with carrier concentration $n$, electron charge $e$, effective scattering time $\tau^* = \tau / (1-\text{i}\omega\tau)$ and scattering time $\tau$, effective mass $m$ and cyclotron frequency $\omega_c=eB/m$.\\

	We first briefly review the well established free carrier response of a system without a TME contribution (for an extensive review see \cite{1970Furdyna}), in order to clearly work out the difference in the magneto-optical response if a TME term is present. We consider a stratified three layer system in which the active film is encapsulated between two layers of vacuum in Faraday geometry. The permittivity in the second (active) layer is deliberately set to 1, which yields the contribution of free carriers only, neglecting any lattice background. We set the effective mass at $m = 0.1 m_0$, which is on the order of magnitude for typical TI materials \cite{Armitage_samples}, $m_0$ as free electron mass. Similarly, the choice for the scattering time is motivated by the reported experimental values of the carrier mobility $\mu$ in HgTe samples (high $10^5\,$cm$^2$/Vs)\cite{2009Roth,2011Bruene,2021Mahler} and Bi2Se3 samples ($3800\,$cm$^2$/Vs)\cite{Armitage_samples}, using the relationship $\tau = \mu\, m/e$.\\
	
\begin{figure}        
		\includegraphics[width=\linewidth]{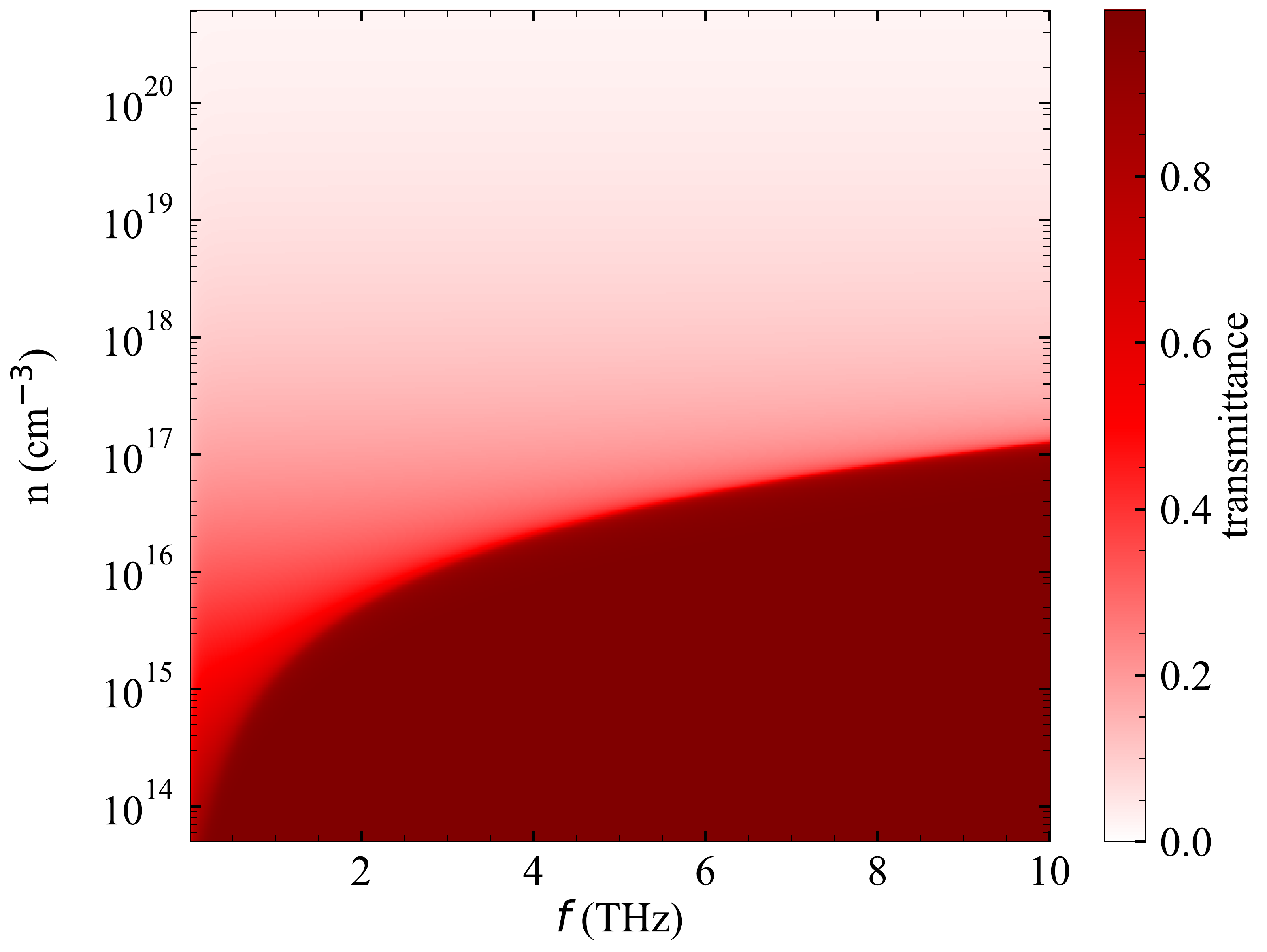}
		\caption{Normalized transmission through the first interface of the three layer systems dependent from multiple orders of magnitude of carrier concentrations in the second layer and the frequency of the incident electromagnetic wave. The normalization is relative to the incident power density in the first layer for each data point.}
		\label{figTheory:plasmaEdge}
	\end{figure}

	Starting without external magnetic field, Fig. \ref{figTheory:plasmaEdge} shows the evolution of the plasma edge, described by $\omega_p~=~\sqrt{\frac{ne^2}{m\, \varepsilon_0\,\varepsilon_{r}}}$ \cite{1970Furdyna}. The smeared-out edge at low frequencies corresponds to the low-frequency limit of the Drude approximation ($\omega\tau << 1$). In this regime, $\varepsilon(\omega)$ is purely imaginary, which results in finite transmission through the interface \footnote{The transmission through a layer of finite thickness generally would depend on the layer thickness, which is ignored here because of thin-film approximation}. At higher frequencies the edge becomes sharper as it approaches the high-frequency limit ($\omega\tau >> 1$), for which $\varepsilon(\omega)$ takes real values. In this limit, two regions arise: For $\omega<\omega_p$ the dielectric function is real valued negative and accordingly the incident EM wave is practically completely reflected. For $\omega>\omega_p$, $\varepsilon(\omega)$ is real valued positive, which results in finite transmission. In the case of our modelled permittivities of $\varepsilon_{r}=1$ of all contributing layers, this results in nearly perfect transmission.
	
\begin{figure}
		\includegraphics[width=\linewidth]{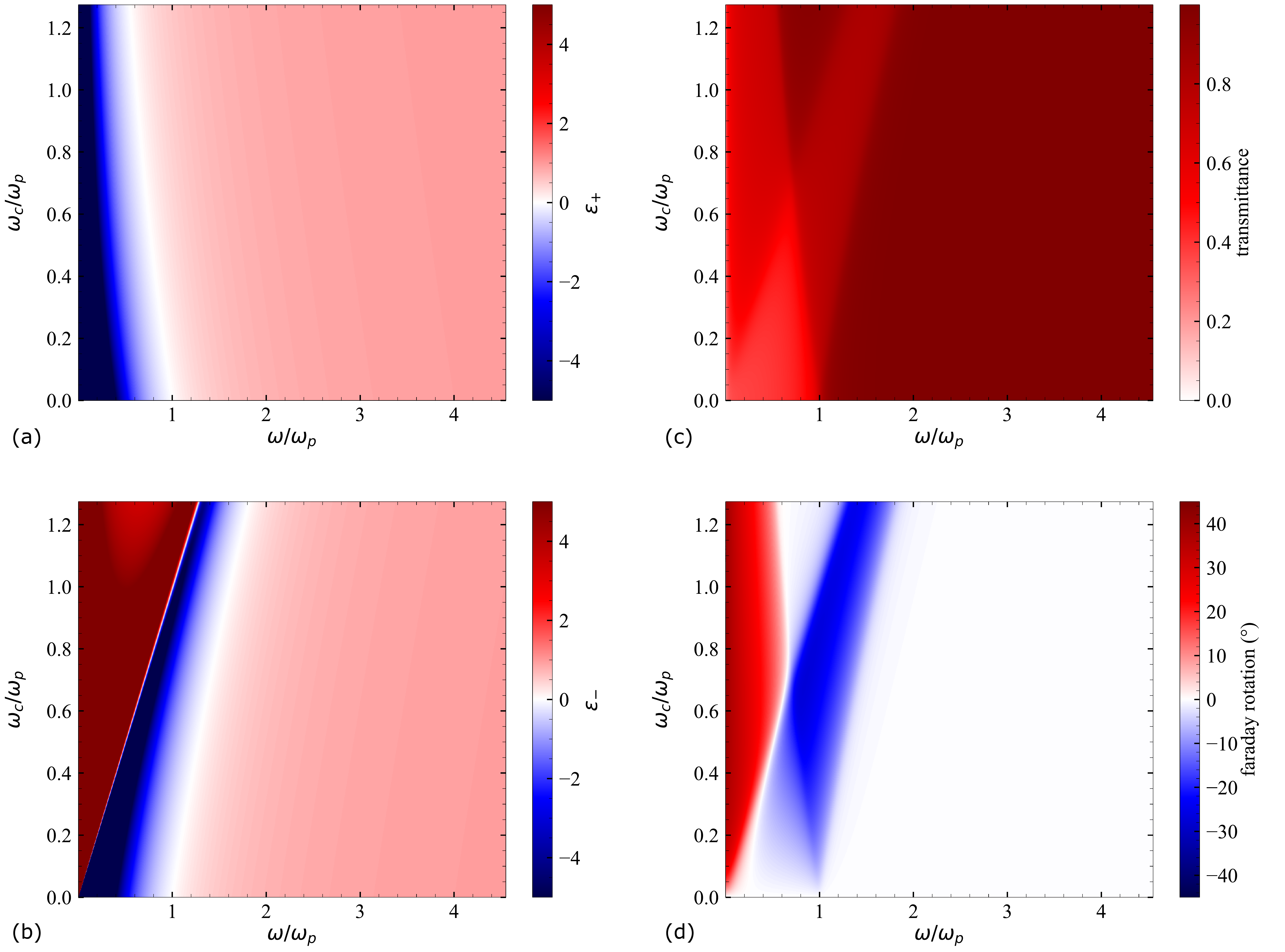}
		\caption{Cyclotron resonance properties for linearly polarized incident EM wave. (a), (b) Real part of dielectric function in circular basis $\varepsilon_{\pm}=\varepsilon_{xx}\pm\text{i}\varepsilon_{xy}$. (c) Normalized transmission through the first interface. (d) Faraday rotation after transmission through the first interface. External magnetic field is along positive $z$-axis. Simulation performed without TME contribution.}
		\label{figTheory:CyclotronResonance}
	\end{figure}

	For the evaluation of the free carrier optical response in external magnetic field, we set the carrier concentration to $n=6 \cdot 10^{15}\,$cm$^{-3}$, which matches the residual bulk carrier concentration in experimentally available high quality TI films~\cite{2007Koenig,2009Roth,2021Mahler}. Fig.~\ref{figTheory:CyclotronResonance} shows the cyclotron resonance properties for the first interface of the three layer system. We first discuss the properties of the cyclotron active circular mode $\varepsilon_{-}$ and the cyclotron inactive mode $\varepsilon_{+}$, displayed in the panels of Fig.~\ref{figTheory:CyclotronResonance}(a) and (b), respectively. The zero-value of the dielectric function traces the evolution of the effective plasma frequency $\omega_p^*$ with external magnetic field. This follows from the definition of $\omega_p$ for zero-field, for which the real part of the dielectric function vanishes for $\omega=\omega_p$.
	
	It is instructive to first consider the lossless case for which $\tau\rightarrow\infty$ (no scattering occurs). In this limit, $\tau^* = \text{i}/\omega$ and as a result $\sigma_{xx}$ is purely imaginary and $\sigma_{xy}$ purely real (see Eqs.~\ref{eqTheory:defConductivityMatrix}). According to Eqs.~\ref{eqTheory:DielectricTensorComponents} this results in a purely real $\varepsilon_{xx}$ and a purely imaginary $\varepsilon_{xy}$, which means in this limit the cyclotron modes $\varepsilon_{\pm}=\varepsilon_{xx}\pm\text{i}\varepsilon_{xy}$ are both purely real.
	
	This implies that transmission of the cyclotron inactive mode through the interface is only possible if $\omega > \omega_p^*$. For the active mode, an additional transmission region occurs for $\omega < \omega_c < \omega_p^*$. From these considerations it is immediately obvious that close to $\omega_c$, for $\omega<\omega_c$ the cyclotron active mode $\varepsilon_{-}$ will dominate the transmission signal and vice versa, for $\omega>\omega_c$ transmission of the cyclotron inactive mode $\varepsilon_{+}$ will. This is also reflected in the total transmission intensity, which we show in Fig.~\ref{figTheory:CyclotronResonance}(c) for finite $\tau$.    
	
	If $\tau$ is finite, the cyclotron resonance is damped and the cyclotron active mode takes the value zero at $\omega_c$. Further, the non-vanishing imaginary part of the dielectric function results in a positive real part of the refractive index even in the regions where Re$(\varepsilon_{\pm})<0$, which enables propagation of the mode. This can be clearly seen in Fig.~\ref{figTheory:CyclotronResonance}(c): In Re$(\varepsilon_{\pm})<0$ regions the transmission is reduced compared to Re$(\varepsilon_{\pm})>0$ regions. The total transmission in these regions, however, does not completely go to zero, because of the non-vanishing imaginary part and partly because the other mode can still be transmitted in this regime. Even if the real part of the dielectric function in both cyclotron modes is negative, there is still finite transmission due to the contribution of the respective imaginary parts of $\varepsilon{\pm}$.
	
	\begin{figure*}
		\includegraphics[width=\linewidth]{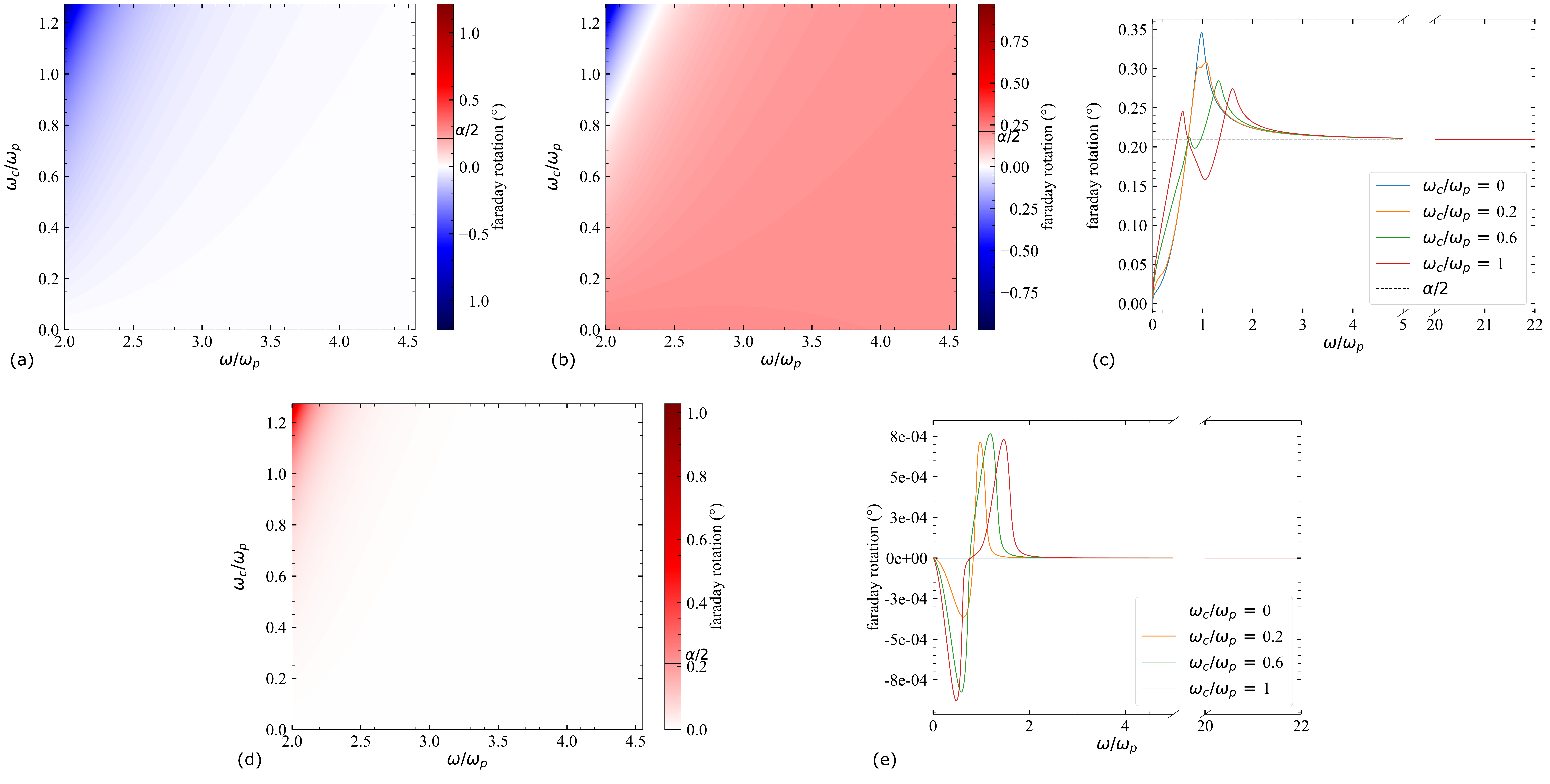}
		\caption{Comparison of Faraday rotation with and without TME contribution. Rotation without (a) (and with (b)) TME contribution after the  first interface. For noticeable difference the resonances are excluded. (c) Difference in rotation between simulation with and without TME contribution over wide frequency range, including resonances, after the first interface. As a guide to the eye the value for $\alpha/2$ is included. (d) Rotation with TME contribution after both interfaces. (e) Difference in rotation between simulation with and without TME contribution over wide frequency range, including resonances, after both interfaces.}
		\label{figTheory:5er_tme}
	\end{figure*}
	
	With these considerations in mind we now inspect carrier induced Faraday rotation upon an incident linearly s-polarized EM wave ($E_{in}=(1, 0)$). From Eq.~\ref{eqTheory:rotation_definition} we obtain for the Faraday angle $\theta_F$:
	
	\begin{equation}
		\theta_F = \text{Re}[\arctan(-t_{sp}/t_{ss})].
		\label{eq:Faraday_angle_w_carriers}
	\end{equation}
	
	Without TME contribution, Eqs.~\ref{eqTheory:TransmittivityWithCarriers} reduce upon transmission through the first interface to
	
	\begin{align}
		&t_{ss} = 2\,(1+\gamma_{xx})\\
		&t_{sp} = -2\,\gamma_{xy}
	\end{align}
	
	with
	
	\begin{align}
		&\gamma_{xx} = \frac{1}{2}\left(\sqrt{\varepsilon_{+}} + \sqrt{\varepsilon_{-}}\right)\\
		&\gamma_{xy} = \frac{\text{i}}{2}\left(-\sqrt{\varepsilon_{+}} + \sqrt{\varepsilon_{-}}\right)	
	\end{align}
	
	The resulting Faraday rotation is shown in Fig.~\ref{figTheory:CyclotronResonance}(d). It resembles the different transmission regions of the respective cyclotron modes and results in huge rotation signals that approach $\pm 45^{\circ}$ in the vicinity of $\omega_c$ for the lossless case. Comparing the resulting rotation angles with the expected rotation from the TME contribution, which is of order $\alpha/2$, it immediately follows that for an unambiguous assignment of the signal to a TME contribution one has to measure far above the cyclotron resonance frequency. To quantify this further, we plot out the respective region of Fig.~\ref{figTheory:CyclotronResonance}(d) on more useful scale for comparison to TME induced Faraday rotation in Fig.~\ref{figTheory:5er_tme}(a). From this plot it becomes clear that the free carrier induced Faraday rotation signal significantly contributes on the scale of the magnitude of the TME induced rotation.
	
	We finally turn to the discussion of the impact of the TME contribution on the magneto-optical response in the presence of itinerant carriers. We keep all parameters but the P$_3$ value of the active layer constant, which we hence set to P$_{3,b}=\frac{1}{2}$, corresponding to the scenario of a 3D TI film. Figs.~\ref{figTheory:5er_tme}(b) and (c) summarize the resulting TME-induced Faraday rotation upon passing through the first interface of the stack. For detection frequencies $\omega$ in the vicinity of $\omega_c$ the Faraday rotation is huge and completely dominated by the free (i.~e.~bulk) carrier response. The presence of the TME contribution modifies the overall response slightly, but the difference is very small. An experimental distinction between the case with and without TME contribution would require accurate knowledge of the the sample properties, which is most probably beyond realistic scenarios. Most importantly, there is no unique feature an experimentalist could look for, apart from some minor magneto-optical response effects that depend on sample details.
	
	At large frequencies $\omega>>\omega_c$, the result of the clean dielectric limit is reproduced and the Faraday rotation angle approaches $\alpha/2$ due to the fact that we set $\varepsilon_r=1$ for the purposes of this discussion. For the more realistic scenario of $\varepsilon_r \neq 1$, the Faraday rotation $\theta_F$ will approach the value for the previously derived scenario of the clean dielectric limit, taking continuous values in the range $\alpha/2 > \theta_F > 0$.
	The total Faraday rotation upon passing through the entire stack is depicted in Figs.~\ref{figTheory:5er_tme}(d) and (e). At this point we recognize an interesting difference with the clean dielectric limit. For $\omega>>\omega_c$ the TME induced contributions to $\theta_F$ again cancel out. In contrast, there is now a net $\theta_F$ resulting from the TME in the spectral region close to $\omega_c$. This stems from the fact that the two interfaces are no longer completely antisymmetric due to the presence of the perpendicular magnetic field and its impact on $\underline{\varepsilon}$. This net signal is, however, very small (order 10$^{-4}$$^\circ$) and certainly beyond the capabilities of contemporary FIR polarimeters.
	
	Summarizing, just like for the case of the clean dielectric limit, the resulting TME induced $\theta_{F}$ is neither quantized nor provides a unique and unambiguously assignable experimental feature. In terms of magnitude of the rotation, the net signal is very small. A clear designation of its origin will require an accurate referencing to the situation without TME contribution, which is in our assessment practically not feasible.

	\subsection{AC Quantum Hall Effect} \label{ssec:idealCase}
	
	Finally, we discuss how the AC Quantum Hall effect is different from a TME response and how the above findings line up with the fact that there are reports of an experimental observation of a universal topological Faraday effect. To this end, we first revisit the Faraday effect arising from 2D (quantum well) systems in general, which is most easily done in circular basis for the transmittivity $t_{\pm}=t_x \pm it_y$. For sake of simplicity and direct comparison, we again consider the geometry discussed in Fig.~\ref{figTheory:rotation_dependency_permittivity_mismatch} with the 2D active layer sandwiched between two layers of vacuum. Under normal incidence, the Fresnel coefficients of the 2D system then take the form~\cite{2011Szkopek} 
	\begin{equation}
		t_{\pm}=\frac{2}{2+Z_0\sigma_{\pm}}
		\label{eq:TransmissionQW}
	\end{equation}
	in which $Z_0$ is the free space impedance and $\sigma_{\pm}=\sigma_{xx}\pm i \sigma_{xy}$ the sheet conductivity. The latter is again a function of the frequency $\omega$. The decisive difference with respect to the previous derivation is that this formula explicitly assumes the 2D situation, which means there is no bulk and accordingly also no associated bulk magneto-electric contribution. The induced Faraday angle is hence given in the usual fashion by
	\begin{equation}
		\tan\theta_F = \frac{t_y}{t_x} = -i \left( \frac{t_+ - t_-}{t_+ + t_-} \right) = -i \left( \frac{\frac{1}{t_-}-\frac{1}{t_+}}{\frac{1}{t_-}+\frac{1}{t_+}}  \right)
		\label{eq:Faraday_QW}
	\end{equation}
	Substituting Eq.~\ref{eq:TransmissionQW} into Eq.~\ref{eq:Faraday_QW} we obtain
	\begin{align}
		\tan\theta_F &= -i \frac{Z_0 (\sigma_- - \sigma_+)}{4 + Z_0(\sigma_- + \sigma_+)} \\
		&=-\frac{Z_0 \sigma_{xy}}{2+Z_0 \sigma_{xx}}
		\label{eq:QuantumFaradayEffect}
	\end{align}
	
	This is a general result and does not include any particular approximations. For any realistic experimental scenario one will of course have to account for the presence of the dielectric contribution of the substrate and potentially the influence of a metal gate (if present). The former is straightforward and will just enter the substrate index of refraction into the denominator of Eq.~\ref{eq:QuantumFaradayEffect}. The latter is more critical and crucially depends on the gate details. For the sake of clarity of the argument we neglect these influences here. 
	
	In the DC Quantum Hall regime, we find the conductances $\sigma_{xx}=0$ and $\sigma_{xy}=\nu \cdot \frac{e^2}{h}$, with $\nu = 1,2,3,\dots$, in which $e$ is the bare electron charge and $h$ the Planck constant. In the low frequency limit $\underline{\sigma}(\omega) \rightarrow \underline{\sigma}(\omega = 0)$, and the Faraday effect is merely an optical probe of the AC Quantum Hall effect. Entering these approximations into Eq.~\ref{eq:QuantumFaradayEffect}, we finally obtain
	\begin{equation}
		\tan\theta_F = - \frac{\nu Z_0 e^2}{2h} = - \nu \alpha.
		\label{eq:ACQuantumHallEffect}
	\end{equation}
	This result was first predicted in the seminal work of Volkov and Mikhailov~\cite{1985Volkov} and resembles what is generally referred to as the AC Quantum Hall Effect. Two conclusions arise immediately upon comparing to the TME. First, even in the hypothetical limit of very clean systems for which the approximation of the clean dielectric limit holds, the observation of a Faraday rotation of integral multiples of $\alpha$ is not unequivocal evidence of the existence of a TME without further experimental insight \footnote{This statement is equally true if one takes the impact of the substrate into account, as has been done by various authors. The procedure changes the absolute value of the Faraday angle the overall sample stack yields, but otherwise the line of argumentation is unchanged.}. The second conclusion arises upon inspecting the employed geometry in the published literature on the TFE. All these works were done in clean Faraday geometry. For the TME this resembles the situation discussed in Fig.~\ref{figTheory:rotation_dependency_permittivity_mismatch}. From the considerations of the previous sections it is clear that a TME can only induce negligibly small Faraday angles and that in this geometry only an AC Hall contribution can be a potential candidate for the origin of the observed signal, as was already pointed out by Beenakker~\cite{2016Beenakker}.
	
	Apart from this, a few comments on the experimental situation must be made. Despite the fundamental character of the quantum hall effect, there is remarkably little work on the dynamical (optical) Hall conductivity $\sigma_{xy}(\omega)$, and even more so in the THz frequency region. The existing experimental work in the the latter regime demonstrates plateau- and steplike features in the Faraday rotation angle, but these are not quantized in units of $\alpha$ and are clearly superimposed on the spectrally close cyclotron resonance \cite{2010Ikebe, 2015Stier}. The situation is different in the microwave spectral region, for which the probing radiation is many (order 10) linewidths below cyclotron resonance, but also here the approximation $\underline{\sigma}(\omega) \rightarrow \underline{\sigma}(\omega = 0)$ does not hold \cite{1986Kuchar, 1993Engel} and the dynamic scaling behaviour of $\underline{\sigma}$ is subject to debate \cite{1993Polyakov, 1996Lee, 2002Hohls}.
	
	The physical origins of the AC QHE in 2D electron systems and the TME in 3D TIs are obviously very different. In particular, the notion of a bulk bears no relevance in the former case. It is, however, surprising that the alleged agreement with the low frequency limit appears to be so much better in 3D TI systems. In practice, the currently available samples are all in the thin film limit for optical measurements and a distinction between AC QHE and TME cannot be acquired in a methodological clean fashion from the optical data alone. The main difference between the TME and the AC QHE is in the geometry of the magnetic field, with which the surface states are gapped out \cite{2013Bernevig}. In a Faraday geometry of a thin film 3D TI, no net Faraday rotation will result from the TME, even if the TME is present. On a more fundamental level the problem is even deeper. For the Faraday geometry employed in the existing experimental publications \cite{2016Armitage,2016Pimenov,2016Okada}, no uniform magneto-electric polarization $P_3$ can be formally defined \cite{2013Bernevig}. If the notion of a TME is then still useful on a conceptual level remains at least doubtful.    
	
	\section{Conclusion}
	We have shown that there is no universally quantized TFE in a 3D TI for which a TME is present. The experimentally observable Faraday rotation angle depends on the exact interplay of TME polarization and non-topologically induced impedance mismatch, and will generally be very small. The vacuum angle of $\alpha/2$ represents the upper boundary and will rapidly approach zero as the regular impedance mismatch sets in. We emphasize that it is also not simply additive as was recently assumed in the experimental analysis of the magneto-optical response of ferromagnetic anomalous quantum Hall samples~\cite{2022Tokura}, for which the TFE contribution actually is vanishingly small. Overall, the real experimental situation is such that there is no unambiguously clear assignable TME signature in the magneto-optical response in the same spirit as a quantized Hall conductance in a DC transport experiment. The decomposition of the net Faraday signal rather requires further detailed input of the material properties.
	
	Methodologically, the TFE and the AC Quantum Hall effect are very difficult to disentangle as both provide a magneto-optical response only at the interfaces. This is particularly true for thin films, in which THz experiments sample the response of top and bottom surfaces of the TI active material simultaneously. The last constraint may be relaxed for thicker layers~\cite{2010Maciejko}, which also allow for the resolution of reflexes coming from different surfaces in time-domain optical experiments. 
	
	For the existing experimental reports on the observation of a TFE, it is certainly possible that the signal stems from the optical response of a DC quantized Hall conductance, but from this one cannot infer that a TME is also present. The TME and the Hall conductance are evidently closely related effects in 3D TIs, but the latter does not necessitate the presence of the former. A noticeable fact is that the key experimental signatures were all observed at high (several Tesla) magnetic fields~\cite{2016Armitage,2016Pimenov,2016Okada}. For the TME in a 3D TI this should not be necessary. It is only required to break time-reversal symmetry, which a small magnetic field also achieves. This raises a fundamental question: At these high magnetic fields, is the system under investigation still a 3D TI or has the magnetic field driven the system already into the quantum Hall state, for which the edge channels may well be composed of bulk states? The Faraday rotation signal of a thin film sample will not be able to distinguish between the two situations, as we show in the AC quantum Hall section.
	
	We conclude that from the existing data no clear answer can be given to this question. It will require further work, both experimental and theoretical, to clearly establish where and how the this transition occurs and how the presence of the TME can be unambiguously pinned down in a non-magnetic sample. This is notably different for ferromagnetic materials in the quantum anomalous Hall regime. Here, the scaling behavior of the flow diagram provides a clear experimental signature for the presence or absence of a TME~\cite{2011Nomura,2017Grauer,2021Fijalkowski}~\footnote{In magnetic 3D TIs the transition from the $\nu=+1$ to the $\nu=-1$ (in units of $e^2/h$) surface transport conductance (or vice versa) can be experimentally tuned by flipping the direction of the magnetization. This is not directly possible in non-magnetic 3D TIs and as a result the scaling behavior of the flow diagram does not provide the same experimental access in non-magnetic materials as it does for ferromagnetic samples.}. It is desirable to establish equally clear experimental fingerprints also for the case of realistic non-magnetic TI materials. 
	
	\acknowledgements{We thank R. Thomale for useful discussions. The authors acknowledge support by the SFB1170 (DFG project ID 258499086) and the W\"urzburg-Dresden Cluster of Excellence on Complexity and Topology in Quantum Matter (EXC 2147, DFG project ID 39085490).}
	
	\clearpage
	\appendix
	\section{Derivation of transmission matrix elements} \label{appsec:derivation_transmission_matrix}
\begin{figure}[H]
		\begin{center}
			\includegraphics[width=\linewidth]{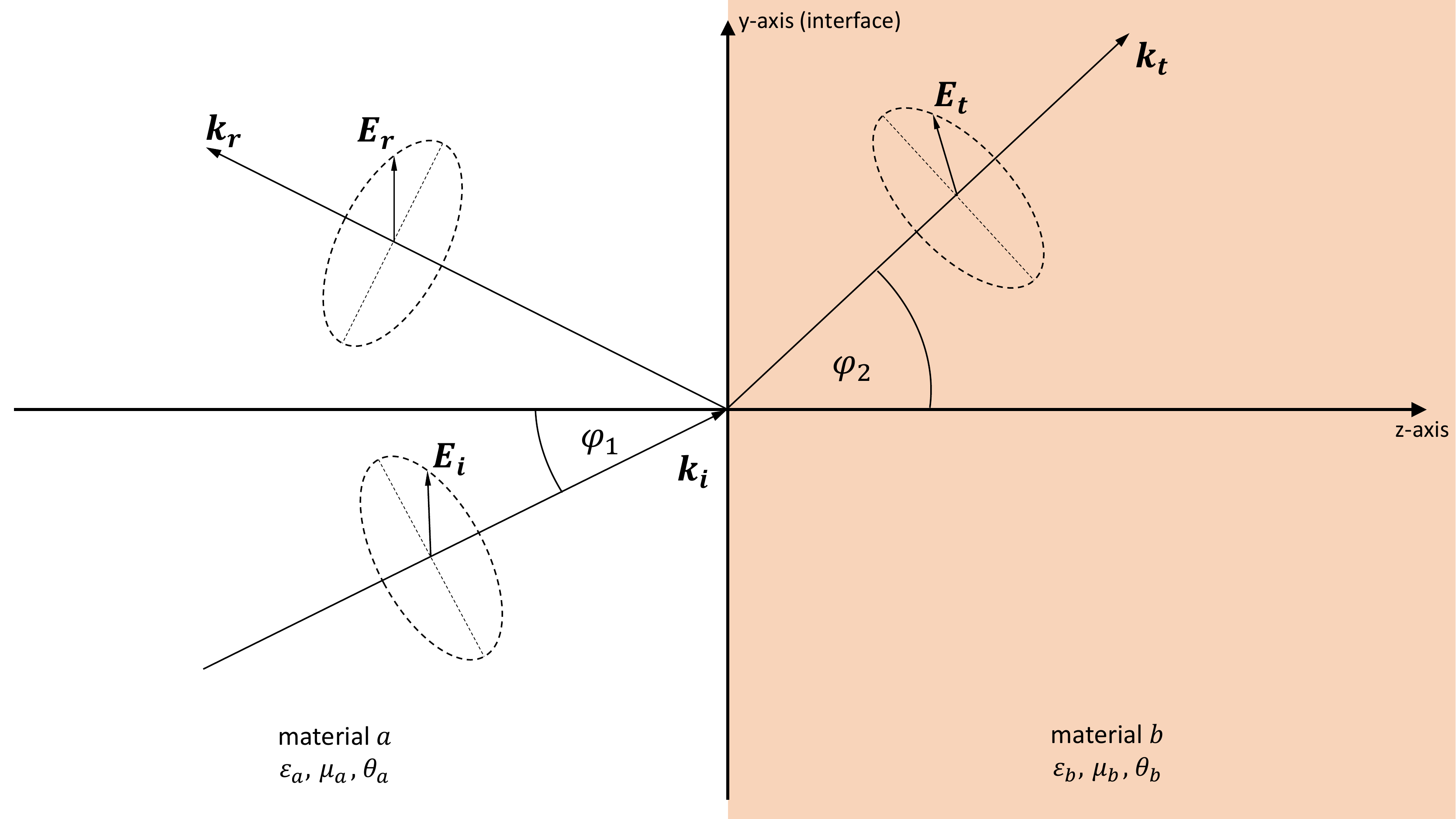}
			\caption{Incident, reflected and transmitted EM wave at an interface, represented by the $x$-$y$-plane, between two materials $a$ and $b$. Only wave vector $\bm{k}$ and the electric field $\bm{E}$ of the EM wave are shown here. The magnetic field $\bm{H}$ can be described by $\bm{k}$ and $\bm{E}$, see main text.}
			\label{figSup:EMwave_on_interface}
		\end{center}
	\end{figure}

	Notation: All bold printed symbols are vectors. If the same symbol is printed regular, it describes the absolute value of this vector. Matrix symbols are underlined.\\\\
	The goal of this derivation is to describe the transmission through an arbitrary interface between two materials $a$ and $b$ and subsequently calculate the polarization state of the electromagnetic wave after transmission. The easiest way to achieve this is using the Jones formalism. A great advantage of this formalism is, that even complex interfaces can be described as matrices. The interaction of the incident light with this interface is easily evaluated by calculating the matrix product of the transmission matrix $\underline{T}$ and the vector of the incident EM wave.
	\begin{align}
		\pmqty{E_{t,\bot}\\E_{t,\|}} &= \underline{T} \pmqty{E_{i,\bot}\\E_{i,\|}}\label{eqSup:defT}\\
		\underline{T} &= \pmqty{t_{ss} & t_{sp}  \\ t_{ps} & t_{pp}}
	\end{align}
	The indices $t$ and $i$ stand for transmitted and incident, $E$ for electric field, $\bot$ and $\|$ for the projections perpendicular and parallel to the incident plane.\\
	To derive the elements of $\underline{T}$, the actual components of incident, reflected and transmitted electric fields of an EM wave at an interface need to be defined first, see fig.\ref{figSup:EMwave_on_interface}. The corresponding components are given in \eqref{eqSup:EMwave_components}.
	\begin{widetext} 
		\begin{align}
			\begin{aligned}
				\bm{E}_i\,&=\, \pmqty{E_{i,\bot}\\E_{i,\|}\cos\varphi_a\\-E_{i,\|}\sin\varphi_a}		&		
				\bm{E}_r\,&=\, \pmqty{E_{r,\bot}\\-E_{r,\|}\cos\varphi_a\\-E_{r,\|}\sin\varphi_a}		&		
				\bm{E}_t\,&=\, \pmqty{E_{t,\bot}\\E_{t,\|}\cos\varphi_b\\-E_{t,\|}\sin\varphi_b}\\ 
				%
				\bm{k}_i\,&=\,k \pmqty{0\\\sin\varphi_a\\\cos\varphi_a}	&		 
				\bm{k}_r\,&=\,k \pmqty{0\\\sin\varphi_a\\-\cos\varphi_a}	&
				\bm{k}_t\,&=\,k_t \pmqty{0\\\sin\varphi_b\\\cos\varphi_b} \\	 
				%
				\bm{k}_i\times\bm{E}_i\,&=\,k\pmqty{-E_{i,\|}\\E_{i,\bot} \cos\varphi_a\\-E_{i,\bot} \sin\varphi_a} 	&	\qquad		
				\bm{k}_r\times\bm{E}_r\,&=\,k\pmqty{-E_{r,\|}\\-E_{r,\bot}\cos\varphi_a\\-E_{r,\bot} \sin\varphi_a} 		&	\qquad	
				\bm{k}_t\times\bm{E}_t\,&=\,k_t \pmqty{-E_{t,\|}\\E_{t,\bot}\cos\varphi_b\\-E_{t,\bot} \sin\varphi_b} \label{eqSup:EMwave_components}
			\end{aligned}
		\end{align}
	\end{widetext}
	Additionally, the continuity conditions for EM fields at an interface will be important:
	\begin{align} 
		\begin{aligned}
			\bm{n}\cdot\bm{D}&=0  \qquad&\bm{n}\cdot\bm{B}=0\\
			\bm{n}\times\bm{E}&=0  \qquad&\bm{n}\times\bm{H}=0 \label{eqSup:continuity_conditions}
		\end{aligned}					
	\end{align}
	Especially the ``$\bm{n}\times$...''-equations are of interest here. But let's take a step back, first. Starting with Maxwell's equation
	\begin{align} 
		\nabla\times\bm{E}=-\mu_r\mu_0\pdv{\bm{H}}{t} \label{eqSup:maxwell_rotE}
	\end{align}
	and the definition of the electric $\bm{E}$ and magnetic field $\bm{H}$
	\begin{align} 
		\bm{E}&=\bm{E_0}\;e^{\text{i}(\bm{k}\bm{x}-\omega t)}\\
		\bm{H}&=\bm{H_0}\;e^{\text{i}(\bm{k}\bm{x}-\omega t)}
	\end{align}
	both derivatives needed in \eqref{eqSup:maxwell_rotE} can be calculated and are given by
	\begin{align} 
		\begin{aligned}
			\nabla\times\bm{E}&=\text{i}\,\bm{k}\times\bm{E}\\
			\pdv{\bm{H}}{t}&=-\text{i}\,\omega\bm{H} \label{eqSup:derivatives_EH}
		\end{aligned}
	\end{align}
	Which results in following relation between $\bm{E}$ and $\bm{H}$, combining \eqref{eqSup:maxwell_rotE} and \eqref{eqSup:derivatives_EH} together:
	\begin{align} 
		\bm{k}\times\bm{E}&=\mu_r\mu_0\,\omega\,\bm{H}\\
		&=\mu_r\mu_0\, c\, k\, \bm{H}\\
		&=\mu_r\mu_0\, \sqrt{\frac{1}{\varepsilon_r\varepsilon_0\mu_r\mu_0}}\,k\,\bm{H}\\
		\nonumber\\
		\Rightarrow\quad\bm{H}&=\sqrt{\frac{\varepsilon_r\varepsilon_0}{\mu_r\mu_0}}\,\frac{1}{k}\;\left(\bm{k}\times\bm{E}\right) \label{eqSup:HpropkcurlE}
	\end{align}
	So, as should be well known, one can describe $\bm{H}$ in terms of $\bm{k}\times\bm{E}$. $\varepsilon_r$ and $\mu_r$ are the relative permittivity and permeability, respectively.\\
	With that, let's focus on \eqref{eqSup:continuity_conditions} again. To fully describe the fields at the interface between two different materials the topological EM constitutive equations need to be taken into account:
	\begin{align} 
		\begin{aligned} 
			\bm{D}=\varepsilon_r\varepsilon_0\bm{E} - 2P_3\alpha\sqrt{\frac{\varepsilon_0}{\mu_0}}\bm{B}\\
			\bm{H}=\frac{1}{\mu_r\mu_0}\bm{B} + 2P_3\alpha\sqrt{\frac{\varepsilon_0}{\mu_0}}\bm{E} \label{eqSup:TME_DH}
		\end{aligned}
	\end{align}
	The TME manifests as collinear coupling between electric and magnetic fields, formally represented in the second summands on the right hand side of \eqref{eqSup:TME_DH}.\\
	With incident, reflected and transmitted fields, from \eqref{eqSup:continuity_conditions} follows
	\begin{align} 
		\bm{n}\times\bm{E_i}+\bm{n}\times\bm{E_r}=\bm{n}\times\bm{E_t}\\
		\bm{n}\times\bm{H_i}+\bm{n}\times\bm{H_r}=\bm{n}\times\bm{H_t}
	\end{align}
	Combining this with \eqref{eqSup:EMwave_components} and \eqref{eqSup:HpropkcurlE} results in
	\begin{widetext} 
		\begin{align}
			\begin{aligned}
				\pmqty{-E_{i,\|}\cos\varphi_a\\E_{i,\bot}\\0}+\pmqty{E_{r,\|}\cos\varphi_a\\E_{r,\bot}\\0}-\pmqty{-E_{t,\|}\cos\varphi_b\\E_{t,\bot}\\0}&=0\\
				\sqrt{\frac{\varepsilon_{r,a}}{\mu_{r,a}}}\left[\pmqty{-E_{i,\bot} \cos\varphi_a\\-E_{i,\|}\\0}+\pmqty{E_{r,\bot}\cos\varphi_a\\-E_{r,\|}\\0}\right]-\sqrt{\frac{\varepsilon_{r,b}}{\mu_{r,b}}}\pmqty{-E_{t,\bot}\cos\varphi_b\\-E_{t,\|}\\0}&\\
				+2\,P_{3,a}\,\alpha\left[\pmqty{-E_{i,\|}\cos\varphi_a\\E_{i,\bot}\\0}+\pmqty{E_{r,\|}\cos\varphi_a\\E_{r,\bot}\\0}\right]-2\,P_{3,b}\,\alpha\pmqty{-E_{t,\|}\cos\varphi_b\\E_{t,\bot}\\0}&=0 \label{eqSup:continuity_conditions_fields_inserted}
			\end{aligned}
		\end{align}
	\end{widetext}
	In the most generalized case $\varepsilon$ is a matrix. For Faraday configuration (external magnetic field parallel to beam propagation) and assuming beam propagation is along $z$-axis (as discussed in the main paper) the resulting dielectric matrix takes the form \cite{1970Furdyna}:
	\begin{align} 
		\underline{\varepsilon}=\pmqty{\varepsilon_{xx}&\varepsilon_{xy}&0 \\ -\varepsilon_{xy}&\varepsilon_{xx}&0 \\ 0&0&\varepsilon_{zz}},
	\end{align} 
	Since the square root of this matrix is needed for \eqref{eqSup:continuity_conditions_fields_inserted}, let's define a new matrix satisfying the condition
	\begin{align}
		\underline{\gamma}\,^2=\underline{\varepsilon}  
	\end{align}
	For given geometry this matrix has the same form as $\varepsilon$
	\begin{align}
		\underline{\gamma}&=\pmqty{\gamma_{xx}&\gamma_{xy}&0\\ -\gamma_{xy}&\gamma_{xx}&0 \\ 0&0&\gamma_{zz}}\label{eqSup:sqrtEpsilon}
	\end{align}
	For further information about the components of $\underline{\gamma}$ refer to Eq.\eqref{eq:gamma_component_definition}.\\
	Performing the matrix product of \eqref{eqSup:sqrtEpsilon} in \eqref{eqSup:continuity_conditions_fields_inserted} leads to
	\begin{widetext}
		\begin{align}
			0=&\left(-E_{i,\|}+E_{r,\|}\right)\cos\varphi_a+E_{t,\|}\cos\varphi_b\label{eqSup:field_components_first}\\
			0=&E_{i,\bot}+E_{r,\bot}-E_{t,\bot}\label{eqSup:field_components_second}\\
			0=&\sqrt{\frac{1}{\mu_{r,a}}}\left[\gamma_{xx,a}\left(E_{r,\bot}-E_{i,\bot}\right)\cos\varphi_a-\gamma_{xy,a}\left(E_{i,\|}+E_{r,\|}\right)\right]
			+\sqrt{\frac{1}{\mu_{r,b}}}\left[\gamma_{xx,b}E_{t,\bot}\cos\varphi_b+\gamma_{xy,b}E_{t,\|}\right]\nonumber\\
			&+2\,P_{3,a}\,\alpha\left(E_{r,\|}-E_{i,\|}\right)\cos\varphi_a
			+2\,P_{3,b}\,\alpha\,E_{t,\|}\cos\varphi_b\label{eqSup:field_components_third}\\
			0=&\sqrt{\frac{1}{\mu_{r,a}}}\left[-\gamma_{xy,a}\left(E_{r,\bot}-E_{i,\bot}\right)\cos\varphi_a-\gamma_{xx,a}\left(E_{i,\|}+E_{r,\|}\right)\right]
			-\sqrt{\frac{1}{\mu_{r,b}}}\left[\gamma_{xy,b}E_{t,\bot}\cos\varphi_b-\gamma_{xx,b}E_{t_\|}\right]\nonumber\\
			&+2\,P_{3,a}\,\alpha\left(E_{i,\bot}+E_{r,\bot}\right)
			-2\,P_{3,b}\,\alpha\,E_{t,\bot}\label{eqSup:field_components_fourth}
		\end{align}
	\end{widetext}
	
	\begin{table}
		\centering
		\resizebox{\columnwidth}{!}{%
			\begin{tabular}{|c||c|c|}
				\hline
				&$E_{r,\bot}$&$E_{r,\|}$\\
				\hline
				\eqref{eqSup:field_components_first}&$0$&$\cos\varphi_a$\\
				\eqref{eqSup:field_components_second}&$1$&$0$\\
				\eqref{eqSup:field_components_third}&$\sqrt{\frac{1}{\mu_{r,a}}}\gamma_{xx,a}\cos\varphi_a$&$-\sqrt{\frac{1}{\mu_{r,a}}}\gamma_{xy,a}+2P_{3,a}\alpha\cos\varphi_a$\\
				\eqref{eqSup:field_components_fourth}&$-\sqrt{\frac{1}{\mu_{r,a}}}\gamma_{xy,a}\cos\varphi_a+2P_{3,a}\alpha$&$-\sqrt{\frac{1}{\mu_{r,a}}}\gamma_{xx,a}$\\\hline
				\hline
				+&$E_{t,\bot}$&$E_{t,\|}$\\
				\hline
				\eqref{eqSup:field_components_first}&$0$&$\cos\varphi_b$\\
				\eqref{eqSup:field_components_second}&$-1$&$0$\\
				\eqref{eqSup:field_components_third}&$\sqrt{\frac{1}{\mu_{r,b}}}\gamma_{xx,b}\cos\varphi_b$&$\sqrt{\frac{1}{\mu_{r,b}}}\gamma_{xy,b}+2P_{3,b}\alpha\cos\varphi_b$\\
				\eqref{eqSup:field_components_fourth}&$-\sqrt{\frac{1}{\mu_{r,b}}}\gamma_{xy,b}\cos\varphi_b-2P_{3,b}\alpha$&$\sqrt{\frac{1}{\mu_{r,b}}}\gamma_{xx,b}$\\\hline
				\hline
				=&$E_{i,\bot}$&$E_{i,\|}$\\
				\hline
				\eqref{eqSup:field_components_first}&$0$&$\cos\varphi_a$\\
				\eqref{eqSup:field_components_second}&$-1$&$0$\\
				\eqref{eqSup:field_components_third}&$\sqrt{\frac{1}{\mu_{r,a}}}\gamma_{xx,a}\cos\varphi_a$&$\sqrt{\frac{1}{\mu_{r,a}}}\gamma_{xy,a}+2P_{3,a}\alpha\cos\varphi_a$\\
				\eqref{eqSup:field_components_fourth}&$-\sqrt{\frac{1}{\mu_{r,a}}}\gamma_{xy,a}\cos\varphi_a-2P_{3,a}\alpha$&$\sqrt{\frac{1}{\mu_{r,a}}}\gamma_{xx,a}$\\\hline
		\end{tabular}}
		\caption{Field component coefficients.}
		\label{tab:field_component_coefficients}
	\end{table}
	
	Evaluating the dependencies of $E_{r,\bot}$, $E_{r,\|}$, $E_{t,\bot}$ and $E_{t,\|}$ from $E_{i,\bot}$ and $E_{i,\|}$ with table~\ref{tab:field_component_coefficients} and Cramer's rule leads directly to the elements of $\underline{T}$, after all \eqref{eqSup:defT} is still valid, which means:
	
	\begin{align} 
		\pmqty{E_{t,\bot}\\E_{t,\|}}=\pmqty{t_{ss}E_{i,\bot}+t_{sp}E_{i,\|}\\t_{ps}E_{i,\bot}+t_{pp}E_{i,\|}}
	\end{align}
	By sorting the resulting dependencies by $E_{i,\bot}$ and $E_{i,\|}$, one can easily identify the matrix elements by comparing coefficients
	\begin{widetext}
		\begin{align}
			&\begin{aligned}
				t_{ss}=&\frac{2}{\Delta}\sqrt{\frac{1}{\mu_{r,a}}}	\left(	\sqrt{\frac{1}{\mu_{r,a}}}\cos\varphi_b\left(\gamma_{xx,a}^2+\gamma_{xy,a}^2\right)
				+\sqrt{\frac{1}{\mu_{r,b}}}\cos\varphi_a\left(\gamma_{xx,a}\gamma_{xx,b}+\gamma_{xy,a}\gamma_{xy,b}\right)
				+2\,\alpha\,\gamma_{xy,a}\cos\varphi_a\cos\varphi_b\left(P_{3,b}-P_{3,a}\right)
				\right)\\
				t_{sp}=&\frac{2}{\Delta}\sqrt{\frac{1}{\mu_{r,a}}}	\left(	\sqrt{\frac{1}{\mu_{r,b}}}\left(\gamma_{xx,b}\gamma_{xy,a}-\gamma_{xx,a}\gamma_{xy,b}\right)
				-2\,\alpha\,\gamma_{xx,a}\cos\varphi_b\left(P_{3,b}-P_{3,a}\right)
				\right)\\
				t_{ps}=&\frac{2}{\Delta}\sqrt{\frac{1}{\mu_{r,a}}}\cos\varphi_a	\left(	\sqrt{\frac{1}{\mu_{r,b}}}\cos\varphi_b\left(\gamma_{xx,a}\gamma_{xy,b}-\gamma_{xx,b}\gamma_{xy,a}\right)
				+2\,\alpha\,\gamma_{xx,a}\left(P_{3,b}-P_{3,a}\right)
				\right)\\
				t_{pp}=&\frac{2}{\Delta}\sqrt{\frac{1}{\mu_{r,a}}}	\left(	\sqrt{\frac{1}{\mu_{r,a}}}\cos\varphi_a\left(\gamma_{xx,a}^2+\gamma_{xy,a}^2\right)
				+\sqrt{\frac{1}{\mu_{r,b}}}\cos\varphi_b\left(\gamma_{xx,a}\gamma_{xx,b}+\gamma_{xy,a}\gamma_{xy,b}\right)
				+2\,\alpha\,\gamma_{xy,a}\left(P_{3,b}-P_{3,a}\right)
				\right)
			\end{aligned}
		\end{align}
		\begin{align}
			\Delta=&		\left(\sqrt{\frac{1}{\mu_{r,a}}}\gamma_{xy,a}\cos\varphi_a+\sqrt{\frac{1}{\mu_{r,b}}}\gamma_{xy,b}\cos\varphi_b+2\,\alpha\,\left(P_{3,b}-P_{3,a}\right)\right)
			\left(\sqrt{\frac{1}{\mu_{r,a}}}\gamma_{xy,a}\frac{\cos\varphi_b}{\cos\varphi_a}+\sqrt{\frac{1}{\mu_{r,b}}}\gamma_{xy,b}+2\,\alpha\,\cos\varphi_b\left(P_{3,b}-P_{3,a}\right)\right)\nonumber\\
			&+	\left(\sqrt{\frac{1}{\mu_{r,a}}}\gamma_{xx,a}\cos\varphi_a+\sqrt{\frac{1}{\mu_{r,b}}}\gamma_{xx,b}\cos\varphi_b\right)
			\left(\sqrt{\frac{1}{\mu_{r,a}}}\gamma_{xx,a}\frac{\cos\varphi_b}{\cos\varphi_a}+\sqrt{\frac{1}{\mu_{r,b}}}\gamma_{xx,b}\right)	\label{eqSup:Delta}
		\end{align}
	\end{widetext}
	Of course, the same method can be employed to get the elements of the reflection matrix $\underline{R}$, if one wanted to investigate the case of reflective geometry. The resulting elements, with $\Delta$ from \eqref{eqSup:Delta}, are:
	\begin{widetext}
		\begin{align}
			\begin{aligned}
				r_{ss}=\frac{1}{\Delta}&\left[		\left(\sqrt{\frac{1}{\mu_{r,a}}}\gamma_{xy,a}\cos\varphi_a-\sqrt{\frac{1}{\mu_{r,b}}}\gamma_{xy,b}\cos\varphi_b-2\,\alpha\,\left(P_{3,b}-P_{3,a}\right)\right)\right.\\
				&\left. \left(\sqrt{\frac{1}{\mu_{r,a}}}\gamma_{xy,a}\frac{\cos\varphi_b}{\cos\varphi_a}+\sqrt{\frac{1}{\mu_{r,b}}}\gamma_{xy,b}+2\,\alpha\,\cos\varphi_b\left(P_{3,b}-P_{3,a}\right)\right)\right.\\
				&\left.	+\left(\sqrt{\frac{1}{\mu_{r,a}}}\gamma_{xx,a}\cos\varphi_a-\sqrt{\frac{1}{\mu_{r,b}}}\gamma_{xx,b}\cos\varphi_b\right)
				\left(\sqrt{\frac{1}{\mu_{r,a}}}\gamma_{xx,a}\frac{\cos\varphi_b}{\cos\varphi_a}+\sqrt{\frac{1}{\mu_{r,b}}}\gamma_{xx,b}\right)\right]\\
				r_{sp}=\frac{2}{\Delta}&\sqrt{\frac{1}{\mu_{r,a}}}	\left(	\sqrt{\frac{1}{\mu_{r,b}}}\left(\gamma_{xx,b}\gamma_{xy,a}-\gamma_{xx,a}\gamma_{xy,b}\right)
				-2\,\alpha\,\gamma_{xx,a}\cos\varphi_b\left(P_{3,b}-P_{3,a}\right)
				\right)\\
				r_{ps}=\frac{2}{\Delta}&\sqrt{\frac{1}{\mu_{r,a}}}\cos\varphi_b	\left(	\sqrt{\frac{1}{\mu_{r,b}}}\cos\varphi_b\left(\gamma_{xx,b}\gamma_{xy,a}-\gamma_{xx,a}\gamma_{xy,b}\right)
				-2\,\alpha\,\gamma_{xx,a}\left(P_{3,b}-P_{3,a}\right)
				\right)\\
				r_{pp}=\frac{1}{\Delta}&\left[		\left(\sqrt{\frac{1}{\mu_{r,a}}}\gamma_{xy,a}\cos\varphi_a+\sqrt{\frac{1}{\mu_{r,b}}}\gamma_{xy,b}\cos\varphi_b+2\,\alpha\,\left(P_{3,b}-P_{3,a}\right)\right)\right.\\
				&\left.	\left(-\sqrt{\frac{1}{\mu_{r,a}}}\gamma_{xy,a}\frac{\cos\varphi_b}{\cos\varphi_a}+\sqrt{\frac{1}{\mu_{r,b}}}\gamma_{xy,b}+2\,\alpha\,\cos\varphi_b\left(P_{3,b}-P_{3,a}\right)\right)\right.\\
				&\left.	-\left(\sqrt{\frac{1}{\mu_{r,a}}}\gamma_{xx,a}\cos\varphi_a+\sqrt{\frac{1}{\mu_{r,b}}}\gamma_{xx,b}\cos\varphi_b\right)
				\left(\sqrt{\frac{1}{\mu_{r,a}}}\gamma_{xx,a}\frac{\cos\varphi_b}{\cos\varphi_a}-\sqrt{\frac{1}{\mu_{r,b}}}\gamma_{xx,b}\right)\right]\\
			\end{aligned}
		\end{align}
	\end{widetext}
	
	We now check for internal consistency with the results obtained for the clean dielectric limit. Therefore, we assume that no carriers are present, so that the whole contribution of the conductivity matrix vanishes and the dielectric function is a diagonal matrix, which only has the entries $\varepsilon_r$, see Eqs.~\eqref{eqTheory:DielectricTensor}~-~\eqref{eqTheory:DielectricTensorComponents}. This reduces to
	\begin{align}
		\gamma_{xx}&=\sqrt{\varepsilon_r}\\
		\gamma_{xy}&=0.
	\end{align}
	
	For the same geometry as in \ref{ssec:clean_dielectric_limit}, we get for one interface (trivial - topological) the transmission matrix elements:
	
	\begin{align}
		t_{ss}=& t_{pp}=\frac{2}{\Delta}\;\sqrt{\frac{\varepsilon_{r,a}}{\mu_{r,a}}}\left(\sqrt{\frac{\varepsilon_{r,a}}{\mu_{r,a}}}+\sqrt{\frac{\varepsilon_{r,b}}{\mu_{r,b}}}\right)\\
		t_{sp}=&-t_{ps}=-\frac{4}{\Delta}\;\sqrt{\frac{\varepsilon_{r,a}}{\mu_{r,a}}}\left(P_{3,b}-P_{3,a}\right)\;\alpha\label{eqTheory:FaradayEpsWithN_comparison_tsp}\\
		\Delta=&\left(\sqrt{\frac{\varepsilon_{r,a}}{\mu_{r,a}}}+\sqrt{\frac{\varepsilon_{r,b}}{\mu_{r,b}}}\right)^2+\left(2\left(P_{3,b}-P_{3,a}\right)\alpha\right)^2
	\end{align}
	which is just a reproduction of matrix elements as \eqref{eqTheory:delta_epsnoN}-\eqref{eqTheory:tss_epsnoN}. Thus, the same rotation angles result and the total rotation is zero, the models are consistent. The same holds for the reflectivity matrix elements, which can be checked in the same fashion.

	\bibliography{biblio}

\end{document}